\documentclass[reprint,aps, prd,letterpaper,showpacs,showkeys,twocolumn,floatfix,nofootinbib,superscriptaddress]{revtex4-1} 

\usepackage{amsmath,amssymb,amsfonts,graphicx}
\usepackage{bm}
\usepackage{slashed}
\usepackage{subfigure}
\usepackage[svgnames]{xcolor}

\newcommand{\vect}[1]{\boldsymbol{#1}}

\begin{document}

\title{Collins fragmentation function within NJL-jet model}

\author{Hrayr~H.~Matevosyan}
\affiliation{CSSM and ARC Centre of Excellence for Particle Physics at the Tera-scale, 
School of Chemistry and Physics, \\
University of Adelaide, Adelaide SA 5005, Australia
\\ http://www.physics.adelaide.edu.au/cssm
}

\author{Anthony~W.~Thomas}
\affiliation{CSSM and ARC Centre of Excellence for Particle Physics at the Tera-scale, 
School of Chemistry and Physics, \\
University of Adelaide, Adelaide SA 5005, Australia
\\ http://www.physics.adelaide.edu.au/cssm
}

\author{Wolfgang Bentz}
\affiliation{Department of Physics, School of Science,  Tokai University, Hiratsuka-shi, Kanagawa 259-1292, Japan
\\ http://www.sp.u-tokai.ac.jp/
}

\begin{abstract}
The Nambu--Jona-Lasinio jet model is extended to accommodate hadronization of a transversely polarized quark in order to explore the Collins effect within a multihadron emission framework. This is accomplished  by calculating the  polarized quark spin flip probabilities after a pseudoscalar hadron emission and the elementary Collins functions. The model is then used to calculate the number densities of the hadrons produced in the polarized quark's decay chain. The full Collins fragmentation function is extracted from the sine modulation of the polarized number densities with respect to the polar angle between the initial quark's spin and hadron's transverse momentum. Two cases are studied here. First, a toy model for elementary Collins function is used to study the features of the transversely polarized quark-jet model. Second, a full model calculation of transverse momentum dependent pion and kaon Collins functions is presented. The remarkable feature of our model is that the 1/2 moments of the favored Collins fragmentation functions are positive and peak at large values of $z$ but decrease and oscillate at small values of $z$. The 1/2 moments of the unfavored Collins functions have comparable magnitude and opposite sign to the favored functions, vanish at large $z$  and peak at small values of $z$. This feature is observed for both the toy model and full calculation and can therefore be attributed to the quark-jet picture of hadronization. Moreover, the  transverse momentum dependencies of the model Collins functions differ significantly from the Gaussian form widely used in the empirical parametrizations.  Finally, a na\"ive  interpretation of the Sch\"afer-Teryaev sum rule is proven not to hold in our model, where the transverse momentum conservation is explicitly enforced. This is attributed to the sizable average transverse momentum of the remnant quark that needs to be accounted for to satisfy the transverse momentum sum-rule.

\end{abstract}

\preprint{ADP-12-20/T787}

\date{\today}                                           

\maketitle

\section{Introduction}

 Deep inelastic scattering (DIS) has been a powerful tool with which to explore the structure of hadrons. Particularly, experiments using the semi-inclusive deep inelastic scattering (SIDIS) have advanced our understanding of the underlying partonic  structure of nucleon in momentum space, allowing access to both the longitudinal and transverse momentum distribution of quarks and gluons. Using polarized probes and/or targets has allowed us to extend these explorations to the spin dependence of the parton distributions. In particular, the naively time-reversal odd Sivers distribution function has been measured to be nonzero \cite{Airapetian:2004tw,Avakian:2003pk, Bradamante:2011xu}. The SIDIS cross sections, both polarized and unpolarized, are a convolution of parton distribution functions with elementary parton-probe scattering amplitudes calculable using perturbative QCD and parton fragmentation functions \cite{Collins:1985ue,*Collins:1989gx}. The complete tree level expressions at leading order have been written in Ref~\cite{Mulders:1995dh}. Thus our detailed understanding of nucleon structure from SIDIS is hinged on our knowledge of fragmentation functions. Moreover, the naive T-odd distributions are convoluted with T-odd fragmentation functions, notably the Collins fragmentation function $H_1^\perp$. The first direct measurements of the Collins mechanism have been performed by the Belle collaboration using hadron pair production in $e^+ e^-$ collisions~\cite{Abe:2005zx,Seidl:2008xc}.
 
  There has been a vast amount of work done to calculate the Collins function within theoretical models. Most notably, the first calculation of a nonvanishing Collins function for pion production within a quark spectator model, using the interference of one-pion-loop amplitudes with the tree level amplitude, was preformed in Refs.~\cite{Bacchetta:2001di,Bacchetta:2002tk}. The resulting Collins fragmentation function appears to be too small to describe the experimental data as shown in Ref.~\cite{Amrath:2005gv}, where the contributions from initial quark rescattering were largely canceled by the quark-pion vertex loop corrections. Later, a similar mechanism was used to calculate the Collins function, through interference of the gluon rescattering and gauge-link correction amplitudes interference with the tree level amplitude Refs.~\cite{Amrath:2005gv,Bacchetta:2007wc, Gamberg:2003eg}. The resulting Collins functions are in generally good agreement with the data, but this approach lacks the ability to produce the unfavored fragmentation functions, including Collins functions. The experimental results from HERMES, COMPASS and JLab are strongly suggesting that the unfavored Collins functions have a similar size and an opposite sign to that of the favored ones~\cite{Airapetian:2004tw,Bradamante:2011xu,Aghasyan:2011ha,Aghasyan:2011gc}.
    
  The NJL-jet model of Refs.~\cite{Ito:2009zc, Matevosyan:2010hh,Matevosyan:2011ey, Matevosyan:2011vj} has been used to calculate the quark fragmentation function within the quark-jet hadronization picture of Field and Feynman~\cite{Field:1976ve,*Field:1977fa}, using the elementary quark-hadron splitting functions calculated within the effective quark model of Nambu and Jona-Lasinio (NJL)~\cite{Nambu:1961tp,*Nambu:1961fr}. The advantage of the model is that it has no free parameters adjusted to fragmentation data and it naturally describes the unfavored fragmentation functions. The use of Monte Carlo (MC) methods in Refs.~\cite{Matevosyan:2011ey, Matevosyan:2011vj} allows for inclusion of hadronic resonances and their decays as well as the transverse momentum dependence of both parton distribution and fragmentation functions. The model was also used to calculate the dihadron fragmentation functions and their evolution in Refs.~\cite{Casey:2012ux,Casey:2012hg}. Here we propose to extend the NJL-jet model by including the spin of the fragmenting quark to calculate the Collins function, both favored and unfavored.  

\begin{figure}[b]
\centering\includegraphics[width=1\columnwidth]{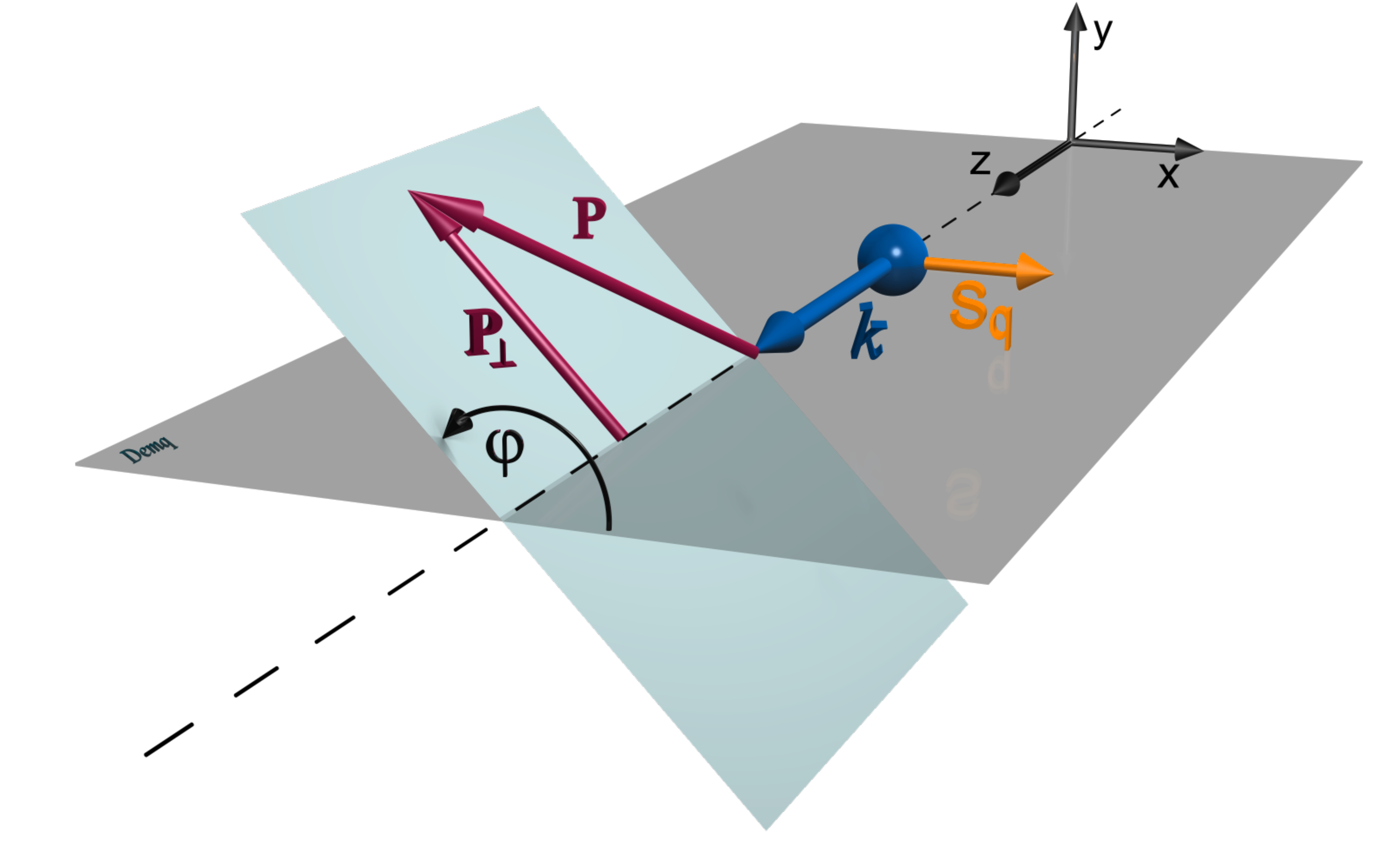}
\caption{Illustration of the three dimensional kinematics of transversely polarized quark fragmentation. 
The fragmenting quark's momentum $\vect{k}$ defines the $z$-axis with its transverse polarization spin vector $\vect{S}_q$ along $x$ axis. The emitted hadron has momentum $P$ with the transverse component $\vect{P_{\perp}}$ with respect to the $z$-axis. The polar angle of hadron's momentum $P$ with respect to the $z x$ plane is denoted by $\varphi$.}
\label{PLOT_POL_QUARK_3D}
\end{figure}

Here we consider the fragmentation of the transversely polarized quark, as depicted schematically in Fig.~\ref{PLOT_POL_QUARK_3D}.
%
 The goal is to calculate the T-odd Collins fragmentation function using the NJL-jet Monte Carlo method of Ref.~\cite{Matevosyan:2011vj}. Following the "Trento Convention" \cite{Bacchetta:2004jz}, the fragmentation function of the transversely polarized quark $q$ to unpolarized hadron $h$ carrying the light-cone momentum fraction $z$ and transverse momentum $\vect{P}_\perp$ with respect to quark's momentum $\vect{k}$ can be expressed as a sum of two terms
\begin{align}
\label{EQ_Dqh}
D_{h/q^{\uparrow}}(z,P_\perp^2,\varphi) &= D_1^{h/q}(z,P_\perp^2)\\ \nonumber
 &+ H_1^{\perp h/q}(z, P_\perp^2) \frac{ (\hat{\vect{k}}  \times \vect{P_\perp}) \cdot \vect{S_q}}{z m_h},
\end{align}
where $\vect{\hat{k}}$ and $\vect{S_q}$ are the momentum and the spin vector of the fragmenting quark, $m_h$ is the mass of the produced hadron. The unpolarized fragmentation function is denoted as $D_1^{h/q}(z,P_\perp^2)$ and $H_1^{\perp h/q}(z, P_\perp^2)$ is the so-called Collins function.  The spin dependent geometric factor multiplying the Collins function in Eq.~(\ref{EQ_Dqh}) can be expressed in terms of the polar angle $\varphi$ of the vector $P$ using the 3-vector components, yielding for the polarized fragmentation function
\begin{align}
\label{EQ_Dqh_SIN}
D_{h/q^{\uparrow}} (z,P_\perp^2,\varphi) &= D_1^{h/q}(z,P_\perp^2)\\ \nonumber
 &- H_1^{\perp h/q}(z, P_\perp^2) \frac{ P_\perp S_q}{z m_h} \sin(\varphi).
\end{align}

 The measure for the above number density is given as
\begin{align}
\label{EQ_Nqh}
&d N_{h/q^{\uparrow}}(z,\vect{P_\perp}) = D_{h/q^{\uparrow}}(z,P_\perp^2,\varphi)\ dz\ d^2\vect{P_\perp} \\ \nonumber
=& \left[D_1^{h/q}(z,P_\perp^2) - H_1^{\perp h/q}(z, P_\perp^2)\frac{P_\perp S_q \sin(\varphi)}{z m_h}\right]dz \frac{d P_\perp^2}{2} d \varphi.
\end{align}

 Integrating over the azimuthal angle $\varphi$ we recover the expression for the number density of the unpolarized quark
 \begin{align}
\label{EQ_Nqh_PP}
d N_{h/q^{\uparrow}}(z,P_\perp^2) & \equiv \int_0^{2\pi} d \varphi\ N_{h/q^{\uparrow}}(z,P_\perp^2,\varphi) \\ \nonumber
&=  \pi D_1^{h/q}(z,P_\perp^2)\ dz\ d P_\perp^2 = d N_{h/q}(z,P_\perp^2).
\end{align}

On the other hand, if we perform an integration over $P_\perp^2$, we arrive at an expression that allows us to extract the $1/2$ moment of the Collins function $H_{1 (h/q)}^{\perp  (1/2)}(z)$
\begin{align}
\label{EQ_Nqh_Phi}
d N_{h/q^{\uparrow}}(z,&\varphi) \equiv D_{h/q^\uparrow}(z,\varphi)\ dz\ d \varphi \\
\nonumber
\equiv &\int_0^{\infty} d P_\perp^2\ N_{h/q^{\uparrow}}(z,P_\perp^2,\varphi) \\
\nonumber
=  &\frac{1}{2\pi}\left[D_1^{h/q}(z)\ - 2H_{1 (h/q)}^{\perp  (1/2)}(z) S_q \sin(\varphi) \right]dz\ d \varphi,
\end{align}
where
\begin{align}
\label{EQ_D1}
D_{1}^{h/q}(z) &\equiv \pi \int_0^{\infty} d P_\perp^2\ D_1^{h/q}(z, P_\perp^2),\\
\label{EQ_H12}
H_{1 (h/q)}^{\perp  (1/2)}(z) &\equiv \pi \int_0^{\infty} d P_\perp^2 \frac{P_\perp}{2z m_h}  H_1^{\perp h/q}(z, P_\perp^2).
\end{align}

 At last, integrating over both $P_\perp^2$  and $\varphi$ in Eq.~(\ref{EQ_Nqh}) we arrive at the integrated number density
\begin{align}
\label{EQ_Nqh_INT}
d N_{h/q^{\uparrow}}(z)\equiv \int_0^{\infty} d P_\perp^2\ \int_0^{2\pi} d \varphi\ N_{h/q^{\uparrow}}(z,P_\perp^2,\varphi) \\ \nonumber
= D_1^{h/q}(z)\ dz.
\end{align}

Our strategy here is to use the NJL-jet model to calculate the number densities $N_{h/q^{\uparrow}}(z,P_\perp^2,\varphi)$ and use the dependence on the azimuthal angle $\varphi$ in Eq.~(\ref{EQ_Nqh}) to extract both $D_1^{h/q}(z, P_\perp^2)$ and $H_1^{\perp h/q}(z, P_\perp^2)$, where the former can be compared with our earlier results in Ref.~\cite{Matevosyan:2011vj} as a cross-check. To achieve this goal, we first expand the NJL-jet model and the associated Monte Carlo framework to accommodate the transverse spin of the fragmenting quark in Sec.~\ref{SEC_NJL-JET-SPIN}. In Sec.~\ref{SEC_DRIVING}, we calculate the elementary unpolarized and Collins fragmentation functions, needed as an input to the NJL-jet model.  In Sec.~\ref{SEC_TOY_MODEL}, we present the results for a simple toy model used as elementary Collins function to demonstrate the distinctive features of the model. In Sec.~\ref{SEC_COLLINS_RES}, we discuss the full model results and present the conclusions and some final remarks in Sec.~\ref{SEC_CONCLUSIONS}.
\section{NJL-jet with transversely polarized quark}
\label{SEC_NJL-JET-SPIN}
\begin{figure}[bp]
\centering\includegraphics[width=1\columnwidth]{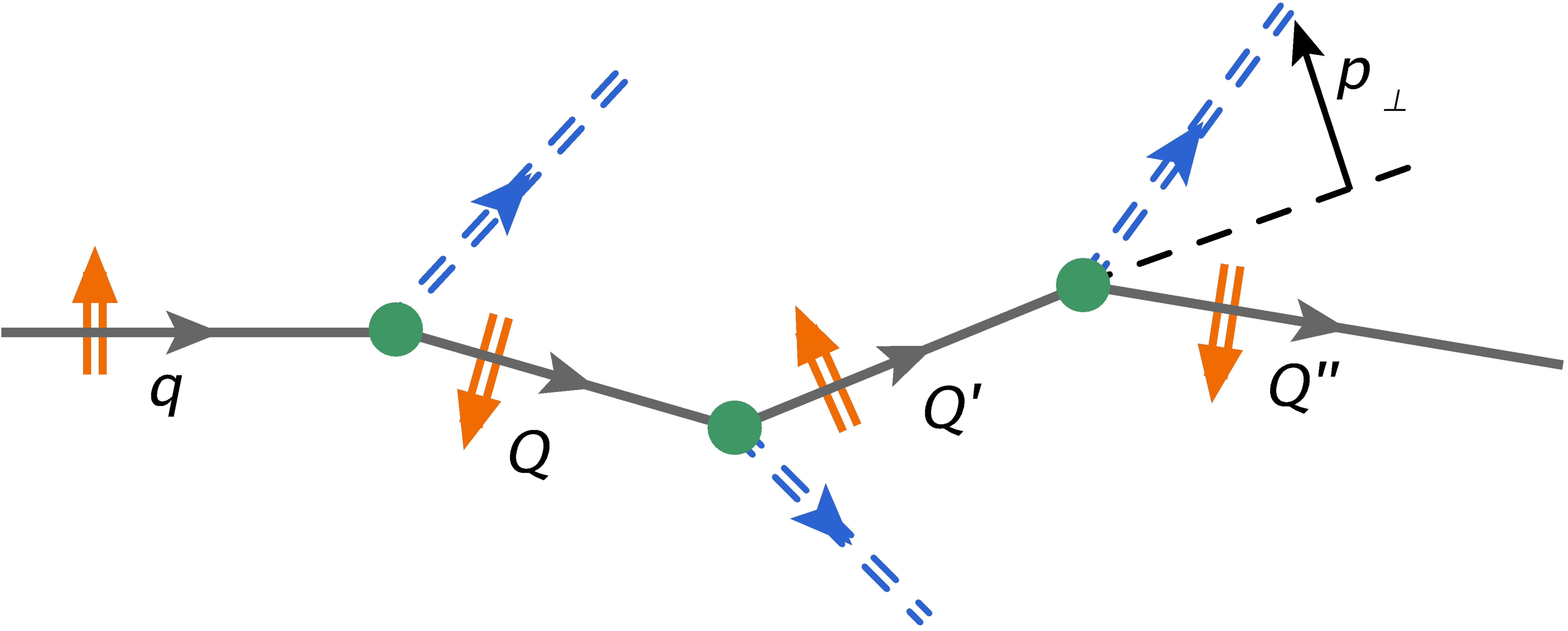}
\caption{NJL-jet model including transverse momentum and quark polarization transfer. Here the orange double-lined arrows schematically indicate the spin direction of the quark in the decay chain.}
\label{PLOT_NJL-JET_TMD}
\end{figure}

 The NJL-jet model employs the quark-jet mechanism to describe the hadronization process.
Here we use the model to describe the fragmentation of the transversely polarized quarks to
unpolarized hadrons, schematically depicted in Fig.~\ref{PLOT_NJL-JET_TMD}.

 An important intricacy arises from the need to keep track of the transverse spin of the quark in the jet as it 
emits hadrons. We first examine the elementary process where a transversely polarized quark emits a single hadron, and we calculate the probability of the final quark's spin being parallel or antiparallel to the original quark's spin direction. In this work we only consider the emissions of pseudoscalar mesons. Further, we make a first order approximation by including in the calculations of the quark spin flip probabilities the elementary hadron emissions only via the tree level diagram depicted in Fig.~\ref{PLOT_FRAG_QUARK}, thus neglecting any T-odd effects. The T-odd effects are considered here to be small relative to the included unpolarized term, though essential for generating the elementary Collins function. This approximation can be easily improved on in the future by also including in the quark spin flip calculation the relevant diagrams that generate the elementary Collins function, such as that discussed in Sec.~\ref{SEC_SUB_COL_SPLITTING}.

 We use the kinematics depicted in Fig.~\ref{PLOT_POL_QUARK_3D}, with the fragmenting quark's momentum defining the $z$ axis. We denote the remnant quark as $Q$ with momentum and spin vectors $l$ and $S_Q$. We use the Dirac spinors of \cite{Cortes:1991ja, Kovchegov:2012ga} to describe the wave functions of transversely polarized quarks
\begin{align}
\label{EQ_SPINOR_T}
U_\chi \equiv \frac{1}{\sqrt{2}}\left[U_{(+z)} + \chi\ U_{(-z)} \right],
\end{align}
where $\chi=\pm 1$ are the eigenvalues of the spin projection onto the $x$ axis and $U_{(\pm z)} $ are the Lepage-Brodsky spinors in helicity basis \cite{Lepage:1980fj,* Brodsky:1997de}. Note that we have a plus sign in front  of  the second term with $\chi$, as there is a sign error in the corresponding expressions of \cite{Cortes:1991ja}.  The spinors $U_\chi$ are both solutions of the Dirac equation and the eigenstates of the Lorentz-covariant spin operator: the Pauli-Lubanski vector
\begin{align}
\label{EQ_PAULI-LUBANSKI}
W_\mu\equiv -\frac{1}{2} \epsilon_{\mu\nu\rho\sigma} S^{\nu\rho}k^\sigma,\\
S^{\nu\rho}\equiv \frac{\imath}{4} \left[ \gamma^\nu, \gamma^\rho \right],
\end{align}
where $\epsilon_{\mu\nu\rho\sigma}$ is the Levi-Civita tensor with convention $\epsilon_{0123} =+1$. It is easy to check explicitly, that
\begin{align}
\label{EQ_SPIN-X_EC}
W_1\ U_\chi = \chi  \frac{m}{2} U_\chi,  \\
(\slashed{k}-m)\ U_\chi =0.
\end{align}

 The normalization of the spinors is
\begin{align}
\label{EQ_SPIN-X_NORM}
\bar{U}_\chi(k,m) U_{\chi'}(k,m)= \delta_{\chi, \chi'} 2m.
\end{align}
  Then the fragmenting quark's spinor is $\Psi_{in}=U_1(k,M_1)$, while the remnant quark's spinor can be described as a superposition of states with spin polarization parallel and antiparallel to the $x$ axis: $\Psi_{out}=a_1\ U_1(l,M_2)+a_{-1}\ U_{-1}(l,M_2)$. Then the relative probabilities of the quark spin flip and nonflip  are determined by $\left|\bar{\Psi}_{out}\gamma^5\Psi_{in}\right|^2$. The spinor matrix elements are given by
\begin{align}
\label{EQ_SPINOR_GA5}
\left| \bar{U}_{\chi '}(l,M_2) \gamma^5  U_{\chi}(k,M_1) \right|^2 =& \\ \nonumber
\delta_{\chi, \chi '} \frac{l_x^2}{1-z}+ \delta_{\chi, -\chi '} & \frac{l_y^2+(M_2-(1-z)M_1)^2}{1-z},
\end{align}
and hence the spin nonflip and spin flip probabilities are proportional to
\begin{align}
\label{EQ_SPIN_FLIP-NON}
| a_1 |^2\sim {l_x^2},\ | a_{-1} |^2 \sim{l_y^2+(M_2-(1-z)M_1)^2}.
\end{align}

 We note, however, that the spinors $U_\chi(k,m)$ become the eigenstates of $W_1$ only if the momentum $k$ has no transverse components: $k_x=k_y=0$. Though the final state quark necessarily has a transverse momentum $\vect{l}_\perp \equiv (l_x,l_y)=-\vect{p_\perp}$, required by the momentum conservation in the transverse plane, it is assumed to be small compared to the light-cone momentum. Thus, in each hadron emission step, we assume the fragmenting quark being in an eigenstate $U_\chi$, and sample the spin flip probabilities according to (\ref{EQ_SPIN_FLIP-NON}). It will be discussed in  Secs.~\ref{SEC_TOY_MODEL} and \ref{SEC_COLLINS_RES} that the results for the Collins function simulations converge rapidly with the number of the hadron emissions in the region of $z\gtrsim0.02$, similar to the case of the unpolarized fragmentation function that was examined in Ref.~\cite{Matevosyan:2011ey}. Thus the failure of this approximation after many hadron emissions, when the light-cone momentum of the quark becomes small and comparable to the transverse component, affects the results only at extremely small values of $z$ and can be neglected.
  
\begin{figure}[tbp]
\centering\includegraphics[width=1.0\columnwidth]{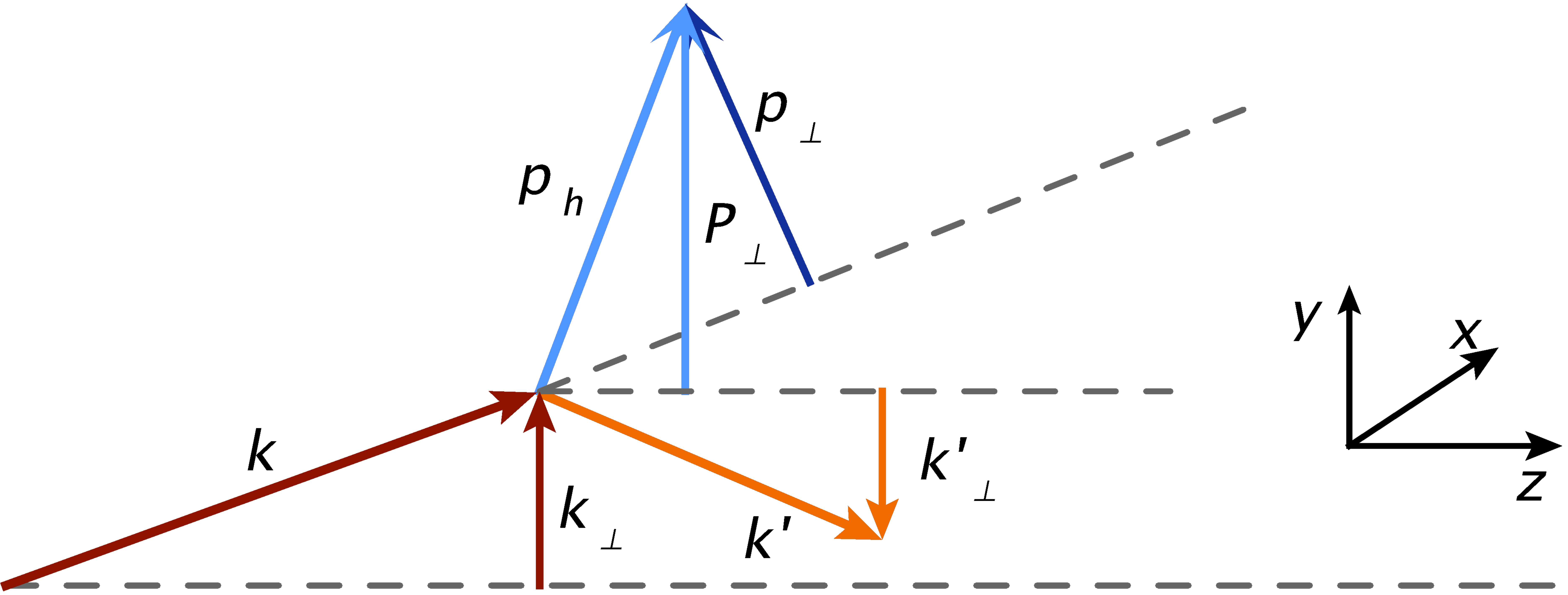}
\caption{Quark elementary fragmentation kinematics, for an arbitrary hadron emission
in the cascade chain. The $z$ axis is defined by the direction of the 3-momentum
of the original parent quark.}
\label{PLOT_SPLITTING_KIN}
\end{figure}

 Using the probabilities of the quark spin flip of Eq.~(\ref{EQ_SPIN_FLIP-NON}) and the elementary polarized
number densities, to be calculated in the next section, we perform a Monte Carlo simulation within the NJL-jet 
framework to calculate the average multiplicities of the produced pseudoscalar mesons (pions and kaons), $N_{h/q^\uparrow}$, as functions of light-cone momentum fraction $z$, transverse momentum square $P_\perp^2$, and polar angle $\varphi$, in the coordinate system defined by the initial quark momentum and spin polarization. Schematically, the elementary hadron emission process is depicted in Fig.~\ref{PLOT_SPLITTING_KIN}, where the vectors $\vect{k}$ and $\vect{k'}$ denote the 3-momentum of an arbitrary quark in the cascade chain before and after hadron emission with transverse components $\vect{k_\perp}$ and $\vect{k'_\perp}$, respectively. The emitted hadron's momentum is labeled by $\vect{p_h}$, where its transverse component with respect to $\vect{k}$ and the $z$ axis is denoted by $\vect{p_\perp}$ and $\vect{P_\perp}$, respectively. $\vect{P_\perp}$ is obtained using the relation $\vect{P_\perp}= \vect{p_\perp} + z\,\vect{k_\perp}$. The recoil transverse momentum of the final quark, $\vect{k'_\perp}$, is calculated from momentum conservation in the transverse plane, namely,
%
\begin{equation}
\label{EQ_PT_QUARK}
\vect{k_\perp}= \vect{P_\perp} + \vect{k'_\perp}.
\end{equation}
  In each hadron emission step we randomly sample the type, the relative light-cone momentum fraction $z$ and transverse momentum and polar angle $\varphi$ (relative to the spin vector of the fragmenting  quark) for the produced hadron according to the relevant elementary number density. We recover and record these quantities in the initial quark's frame, as well as sample the spin direction of the final quark according to (\ref{EQ_SPIN_FLIP-NON}). Then we can extract the number densities corresponding to Eq.~(\ref{EQ_Nqh}) in the usual way \cite{Matevosyan:2011ey,Matevosyan:2011vj}
\begin{widetext}
\begin{align}
\label{EQ_MC_EXTRACT}
D_{h/q^{\uparrow}}(z, P_\perp^2,\varphi)\ \Delta z\ \frac{\Delta P_\perp^2}{2}\ \Delta \varphi
&= \left< N_{q^\uparrow}^h(z,z+\Delta z; P_\perp^2,P_\perp^2  +\Delta P^2; \varphi, \varphi+\Delta \varphi )\right>\\ \nonumber
&\equiv  \frac{ \sum_{N_{Sims}} N_{q^\uparrow}^h(z, z+ \Delta z,P_\perp^2,P_\perp^2+ \Delta P_\perp^2; \varphi, \varphi+\Delta \varphi)} { N_{Sims} },
\end{align}
where $N_{q^\uparrow}^h$ is the number of the hadrons $h$ produced by the quark $q$ that have momentum components laying within the regions specified in its arguments and $N_{Sims}$ is the number of quark decay chain simulations performed.


\section{Elementary Fragmentations}
\label{SEC_DRIVING}
 
 The quark fragmentation functions are defined using the quark-quark correlators~\cite{Collins:1985ue,Collins:1989gx,Bacchetta:2006tn,Bacchetta:2007wc}

\begin{align}
\label{EQ_FRAG_CORR}
\Delta(z, \vect{P}_\perp)=\sum_X \int \frac{d \xi^- d^2 \vect{\xi}_T }{4z (2\pi)^3} e^{\imath k \cdot \xi} & \left< 0 |\ \mathcal{U}^T[\infty_T, \vect{\xi}_T; -\infty^-]\ \mathcal{U}^-[ -\infty^-,\xi^-; \vect{\xi}_T]\ \psi(\xi) | h, X  \right>  \\ \nonumber
\times &\left. \left< h, X |\ \bar{\psi}(0)\  \mathcal{U}^-[ 0^-,-\infty^-; 0_T]\ \mathcal{U}^T[0_T, \infty_T; -\infty^-]  |0 \right> \right |_{\xi^+=0},
\end{align}
\end{widetext}
 where $\psi$ is the quark wave function, and $\mathcal{U}^{(-, T)}$ represents the gauge links along the minus and transverse directions in light-cone coordinates. The details of the gauge link structure can be found in Ref.~\cite{Bacchetta:2006tn}. 
 
  In this section we aim to calculate the elementary unpolarized and Collins fragmentation functions (or splitting functions), corresponding to only a single hadron emission truncation of the relation~(\ref{EQ_FRAG_CORR}). We use the framework of the NJL effective chiral quark model for this purpose. The corresponding unpolarized splitting functions, have been calculated in Refs.~\cite{Ito:2009zc, Matevosyan:2010hh, Matevosyan:2011ey}, where the model parameters were also fixed. In the next subsection we briefly summarize their results and refer the reader to the original articles for the details. For the Collins splitting functions, we adopt the gauge link coupling mechanism to the final quark of Refs.~\cite{Amrath:2005gv,Bacchetta:2007wc, Gamberg:2003eg}, but use the NJL model formalism and parameters. 
 
\subsection{Unpolarized splitting function}
\label{SEC_SUB_UNP_SPLITTING}

 The unpolarized fragmentation function, $D_1^{h/q}(z,\vect{P_\perp})$, is defined as a trace of the quark-quark correlator
\begin{align}
\label{EQ_D1_CORR}
D_1(z,P_\perp^2)=\mathrm{Tr}[\Delta[z,P_\perp]\gamma^+]/2.
\end{align}

\begin{figure}[t]
\centering\includegraphics[width=0.8\columnwidth]{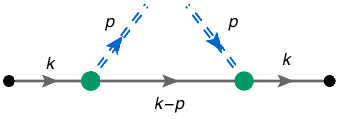}
\caption{Feynman diagram describing the elementary quark to hadron fragmentation functions.}
\label{PLOT_FRAG_QUARK}
\end{figure}

The single hadron emission cut diagram for a quark $q$, to emit a meson $h$, carrying light-cone momentum fraction $z$, and transverse momentum $p_\perp$, is depicted in Fig.~\ref{PLOT_FRAG_QUARK}. In the frame where the fragmenting quark has zero transverse momentum, but a nonzero transverse momentum component $-\vect{p_\perp}/z$ with respect to the direction of the produced hadron~\cite{Collins:1977iv,Ito:2009zc}, the unregularized elementary transverse-momentum-dependent (TMD) fragmentation functions to pseudoscalar mesons are given by
\begin{align}
\label{EQ_QUARK_FRAG_TMD}
d_{1}^{h/q}(z,p_\perp^2)
&= \frac{C_q^h }{16\pi^{3}}\, g_{hqQ}^{2}\, z\ \\\nonumber
&\times  \frac{p_{\perp}^{2} + \left[(z-1)M_{1}+M_{2}\right]^{2}} 
{\left[p_{\perp}^{2}+z(z-1)M_{1}^{2}+zM_{2}^{2}+(1-z)m_{h}^{2}\right]^{2}},
\end{align}
where $M_1$ and $M_2$ denote the masses of fragmenting and remnant quarks. Quark flavor is also indicated by the subscripts $q$ and $Q$, where a meson of type  $h$ has the quark flavor structure $h=q\overline{Q}$ and $m_h$ denotes the meson mass. The corresponding isospin factor and quark-meson coupling constant are labeled by $C_q^h$ and $g_{hqQ}$, respectively, and have been determined  within the NJL model~\cite{Matevosyan:2010hh,Matevosyan:2011vj}.

 In the model, the transverse momentum integrated fragmentation of Eqs.~(\ref{EQ_D1},\ref{EQ_H12}) (as well as the quark-hadron couplings, etc.) require an introduction of a regularization scheme. In this work, the  integrals with two particles in the loop with masses $\mu_1$ and $\mu_2$, carrying light-cone momentum fractions $x$ and $1-x$, respectively, are regularized using the modified Lepage-Brodsky transverse momentum cutoff functions, as described in~\cite{Matevosyan:2011ey}
\begin{align}
\label{EQ_LB_DIP}
G_{12}(p_\perp^2) \equiv \frac{1}{\left[ 1 + (M_{12}^2/\Lambda_{12}^2)^2 \right]^2},
\end{align}
where $M_{12}^2$ is their invariant mass squared. In the frame where the total transverse momentum is zero, $M_{12}^{2} = (\mu_{1}^{2}+p_{\perp}^{2})/x + (\mu_{2}^{2}+p_{\perp}^{2})/(1-x)$. The cutoff,  $\Lambda_{12}$, is 
determined by
\begin{align}
\label{EQ_LB_REG}
M_{12}^2\leq \Lambda_{12}^2 \equiv \left(\sqrt{\Lambda_{3}^{2} + \mu_1^{2}} +\sqrt{\Lambda_{3}^{2} + \mu_2^{2}}\right)^2,
\end{align}
where $\Lambda_3$ is the 3-momentum cutoff, that is fixed in the usual way by
reproducing the experimental pion decay constant. We use a light constituent quark
mass $M=0.4~\mathrm{GeV}$, for which the corresponding 3-momentum cutoff is $\Lambda_3= 0.77~{\rm GeV}$. The strange constituent quark mass of $M_s=0.59~\mathrm{GeV}$ is obtained by reproducing the experimental kaon mass. The model values for the quark-meson coupling constants are determined from the residue at the pole in the quark-antiquark t matrix to be $g_{\pi qQ}=4.24$ and $g_{K qQ}=4.52$.

\subsection{Elementary Collins function}
\label{SEC_SUB_COL_SPLITTING}
\vspace{-0.2cm}
 The Collins function is defined via the trace of the quark-quark correlator
\begin{align}
\label{EQ_COL-FRAG_DEF}
\frac{\epsilon_T^{ij} k_{T j}}{m_h} H_1^\perp(z, P_\perp^2)= \mathrm{Tr}[\Delta(z, \vect{P}_\perp) \imath \sigma^{i-}\gamma_5],
\end{align}
where $m_h$ is the mass of the produced hadron,  $\epsilon_T^{12}=-\epsilon_T^{21}=1$, and $\epsilon_T^{11}=\epsilon_T^{22}=0$ .

\begin{figure}[h]
\centering\includegraphics[width=0.8\columnwidth]{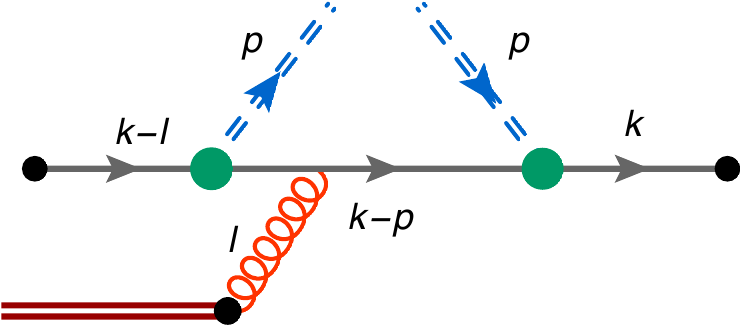}
\caption{Cut diagram describing the elementary quark to hadron Collins fragmentation functions.}
\label{PLOT_FRAG_COL_QUARK}
\end{figure}

 It was shown explicitly, that the simple tree level approximation for the correlator, used to calculate the unpolarized quark splitting function, yields a zero result for the Collins function \cite{Bacchetta:2001di,Amrath:2005gv}. Thus, here we generate the model Collins function through the interference between the tree level hadron emission amplitude and the one with a gauge-link interaction with the final state quark via single gluon exchange, as described in Ref.~\cite{Bacchetta:2007wc}. Note, that we omit the gluon-quark rescattering and gluon loop correction to the quark-hadron vertex, as those nearly cancel each other.  Also, we note that the amplitudes written in the spectator formalism coincide with those in NJL, except for the different quark-hadron coupling constant, the remnant "spectator" being treated as a quark in NJL and the loop integral regularization scheme employed in the NJL-jet. Thus we can simply use the results from Ref.~\cite{Bacchetta:2007wc} with just a few modifications. The relevant diagram for the NJL-jet model is depicted in Fig.~\ref{PLOT_FRAG_COL_QUARK}.

 The result for the elementary Collins function is
\begin{align}
\label{EQ_COL-FRAG_ELEM}
\nonumber
\widetilde{H}_1^\perp(z, p_\perp^2)=& -\frac{2 \alpha_s }{(2 \pi)^4} C_\mathrm{F} \frac{C_q^m}{2}  g_{mqQ}^{2} \frac{m_h}{1-z} \frac{1}{k^2-M_1^2}
\frac{z}{2 p_\perp^2} \\ \nonumber
&\times \{  -I_{34g}\left[(2 z -1) M_1 + M_2\right]  \\ \nonumber
& + I_{2g}\left[2zM_1( k^2 - M_1^2 + m_h^2(1-2/z)) \right. \\ \nonumber
&+(M_2-M_1)( (2z-1)k^2 - m_h^2  \\ 
&\left. + M_2^2 -2z M_1 (M_1+M_2)) \right]  \},
\end{align}
where $\alpha_s$ is the strong coupling constant, $C_\mathrm{F}=4/3$ is the cubic Casimir invariant of the color $SU(3)$ group and the initial quark's total momentum squared, $k^2$, can be expressed as $k^2= {p_\perp^2}/{(z(1-z))}+{M_2^2}/{(1-z)}+{m_h^2}/{z}.$ Since only the imaginary part of the amplitudes contribute to the above expression, the loop integrals for those can be calculated without the need for explicit regularization
\begin{align}
\label{EQ_COL-LOOP_INT}
\nonumber
&I_{2g}= \frac{\pi}{2\sqrt{\lambda( M_2, m_h)}} \ln \left[ \frac{k^2 +M_2^2 -m_h^2 -\sqrt{\lambda( M_2, m_h)}}{k^2 +M_2^2 -m_h^2 +\sqrt{\lambda( M_2, m_h)}} \right], \\ 
&I_{34g} = \pi \ln \left[ \frac{\sqrt{k^2}(1-z)}{M_2} \right],\\ \nonumber
&\lambda(m_1,m_2)= (k^2-(m_1+m_2)^2)(k^2-(m_1-m_2)^2).
\end{align}

 Then the elementary polarized fragmentation function is given as
\begin{align}
\label{EQ_ELEM_Dqh_SIN}
d_{h/q^{\uparrow}} (z,p_\perp^2,\varphi)&= d_1^{h/q}(z,p_\perp^2)- \widetilde{H}_1^{\perp h/q}(z, p_\perp^2) \frac{ p_\perp S_q}{z m_h} \sin(\varphi),
\end{align}
where we use the multiplicative regulator of Eq.~(\ref{EQ_LB_DIP}) for integrals over $p_\perp^2$. The only remaining parameter is the strong coupling in Eq.~(\ref{EQ_COL-FRAG_ELEM}). We consider it as a model parameter, that we fix to the largest value of $\alpha_s=0.444$, that still allows for the positivity bound to be satisfied, namely $d_{h/q^{\uparrow}} \geqslant 0$.  At the next-to-leading order, this value corresponds to a typical hadronic scale of $Q^2=1~\mathrm{GeV^2}$, which is much higher than the one typically used as the NJL-jet model scale, namely $Q_0^2=0.2~\mathrm{GeV^2}$. Such a discrepancy between the scales for the unpolarized and Collins functions is rooted in the model for the Collins function employed here, which would violate the positivity bound if calculated at the typical model scale. A similar issue was encountered in the original work of  Ref.~\cite{Bacchetta:2007wc}, where the value of $\alpha_s=0.2$ was chosen, much smaller than that for the scale of their model set at $Q^2=0.4~\mathrm{GeV}^2$. A completely consistent determination of a single scale for the polarized fragmentation function must involve the QCD evolution of both the TMD unpolarized and Collins functions. This is not possible at present, because the evolution equation for the Collins function is unknown.
 
\begin{figure}[t]
\centering 
\subfigure[] {
\includegraphics[width=0.8\columnwidth]{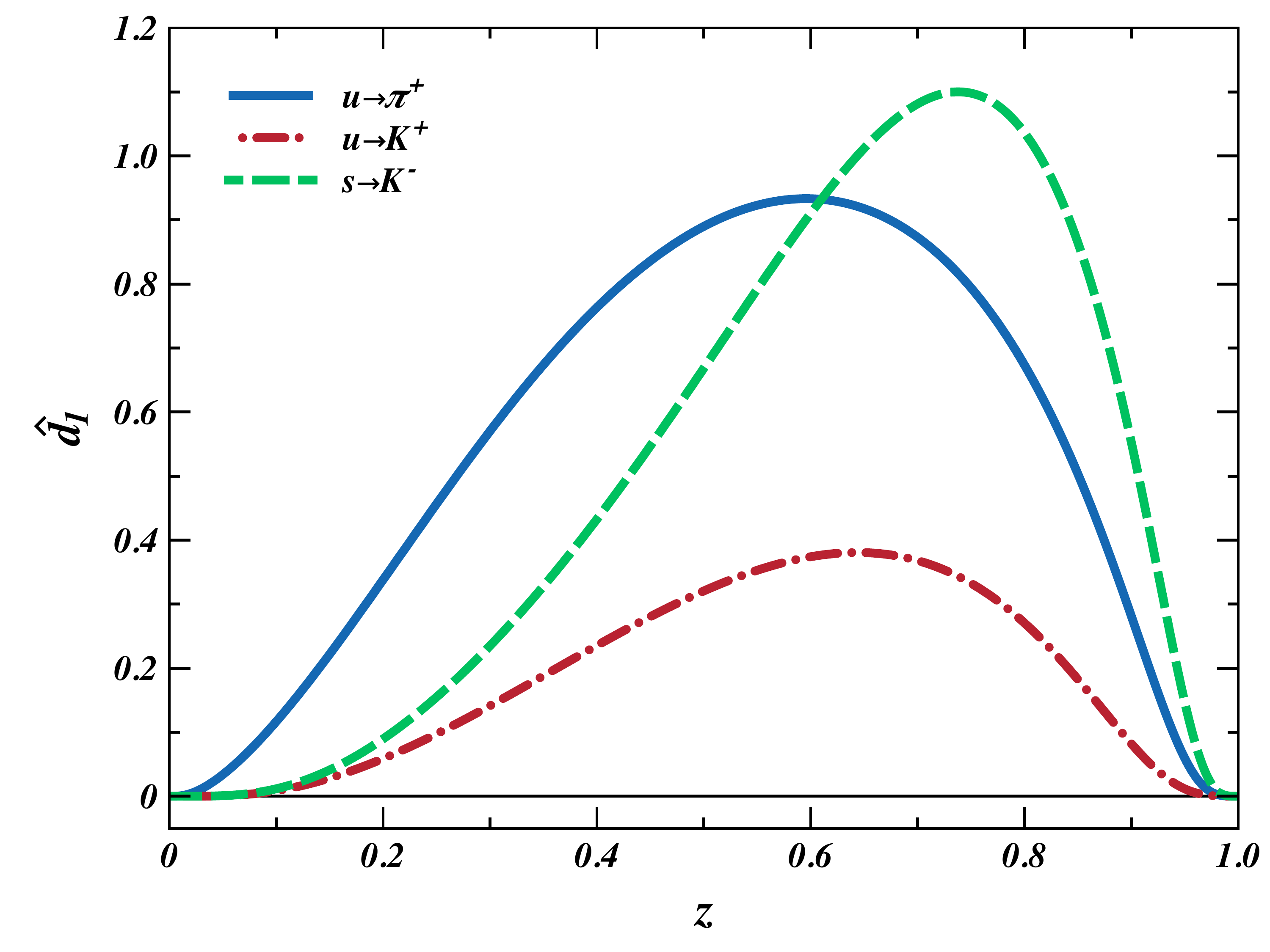}
}
\vspace{0.cm} 
\subfigure[] {
\includegraphics[width=0.8\columnwidth]{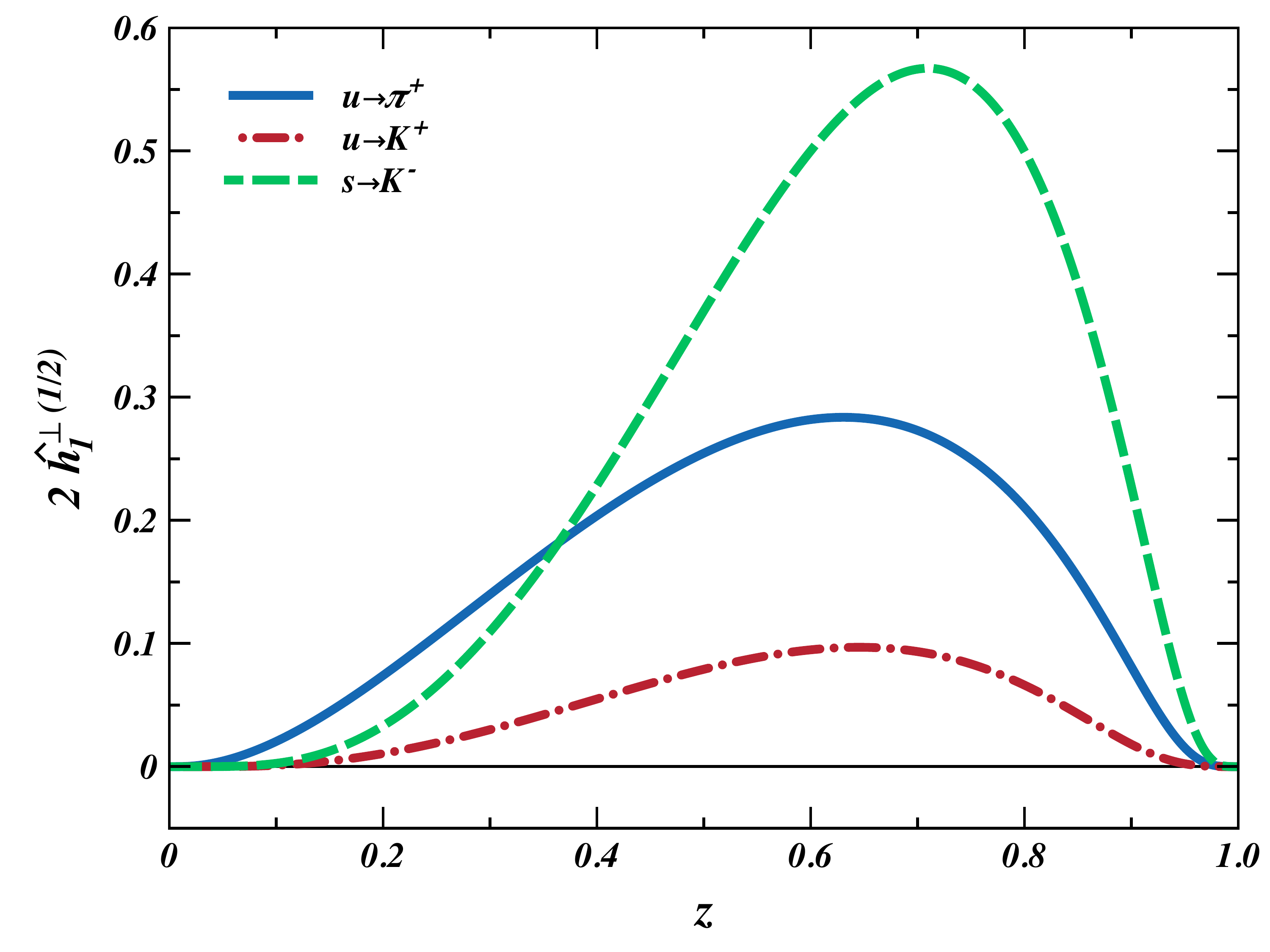}
}
\caption{Elementary renormalized unpolarized fragmentation function, $\hat{d}_1$, (a) and Collins function $1/2$ moment, $2\hat{H}_1^{\perp (1/2)}$, (b) used in the full model calculations of Sec.~\ref{SEC_COLLINS_RES}.}
\label{PLOT_ELEM_FRAG}
\end{figure}

 Within the NJL-jet model, the elementary splitting functions are renormalized such that quark's total probability of emitting a hadron in each step is one: $\sum_h \int dz\ dp_\perp^2/2\ d \varphi\ \hat{d}_{h/q^{\uparrow}} (z,p_\perp^2,\varphi)=1$, the sum is over all hadrons the quark of given flavor can emit directly. These renormalized splittings will be used in the next two sections as input to the Monte Carlo simulations of the quark-jet hadronization process. The integrated renormalized elementary fragmentation functions for the full model calculations of Sec.~\ref{SEC_COLLINS_RES} are depicted in Fig.~\ref{PLOT_ELEM_FRAG}.
 
\section{The Quark-jet effects on Collins function using a Toy Model}
\label{SEC_TOY_MODEL}

 In this section we employ a toy model for the elementary Collins function to explore the general features of the NJL-jet model extended to transversely polarized quark fragmentation, as described in  Sec.~\ref{SEC_NJL-JET-SPIN}. In this toy model we assume that  $-\widetilde{H}_{1 }^{\perp h/q}(z,p_\perp^2)\frac{p_\perp S_q}{m_h z} = 0.1\ d_1^{h/q}(z,p_\perp^2)$. Thus for the elementary number density we simply have
\begin{align}
\label{EQ_MC_DRV_TOY}
d_{h/q^{\uparrow}}^{(toy)} (z,p_\perp^2)=  d_1^{h/q}(z,p_\perp^2)(1+0.1 \sin{\varphi}),
\end{align}
where $d_1^{h/q}(z,p_\perp^2)$ is given by Eq.~(\ref{EQ_QUARK_FRAG_TMD}).
\begin{figure}[h]
\centering
\includegraphics[width=0.9\columnwidth]{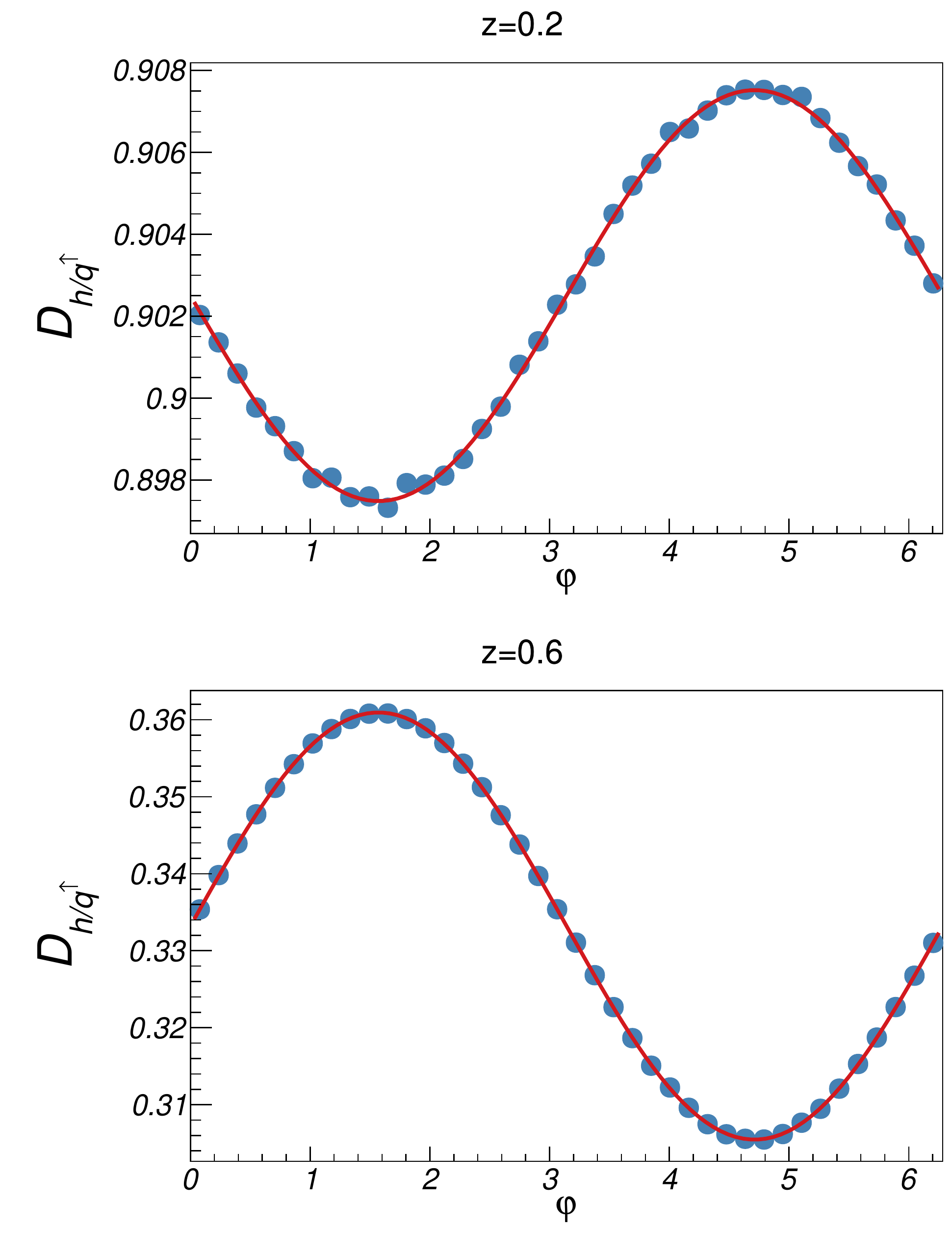}
\caption{The histograms (blue dots) for polarized number density $D_{\pi^0/u^\uparrow}(z,\varphi)$ for two values of $z$ as a function of the azimuthal angle $\varphi$ from NJL-jet framework using a toy model with a fixed number of hadrons emitted in each decay chain $N_{Links}=6$. The minimum-$\chi^2$ fits with a functional form $D_{h/{q^\uparrow}}= c_0+c_1 \sin (\varphi)$  (red lines) are also depicted.}
\label{PLOT_DPOL_TOY_PIP0}
\end{figure}
\begin{figure}[pthb]
\centering 
\subfigure[] {
\includegraphics[width=0.9\columnwidth]{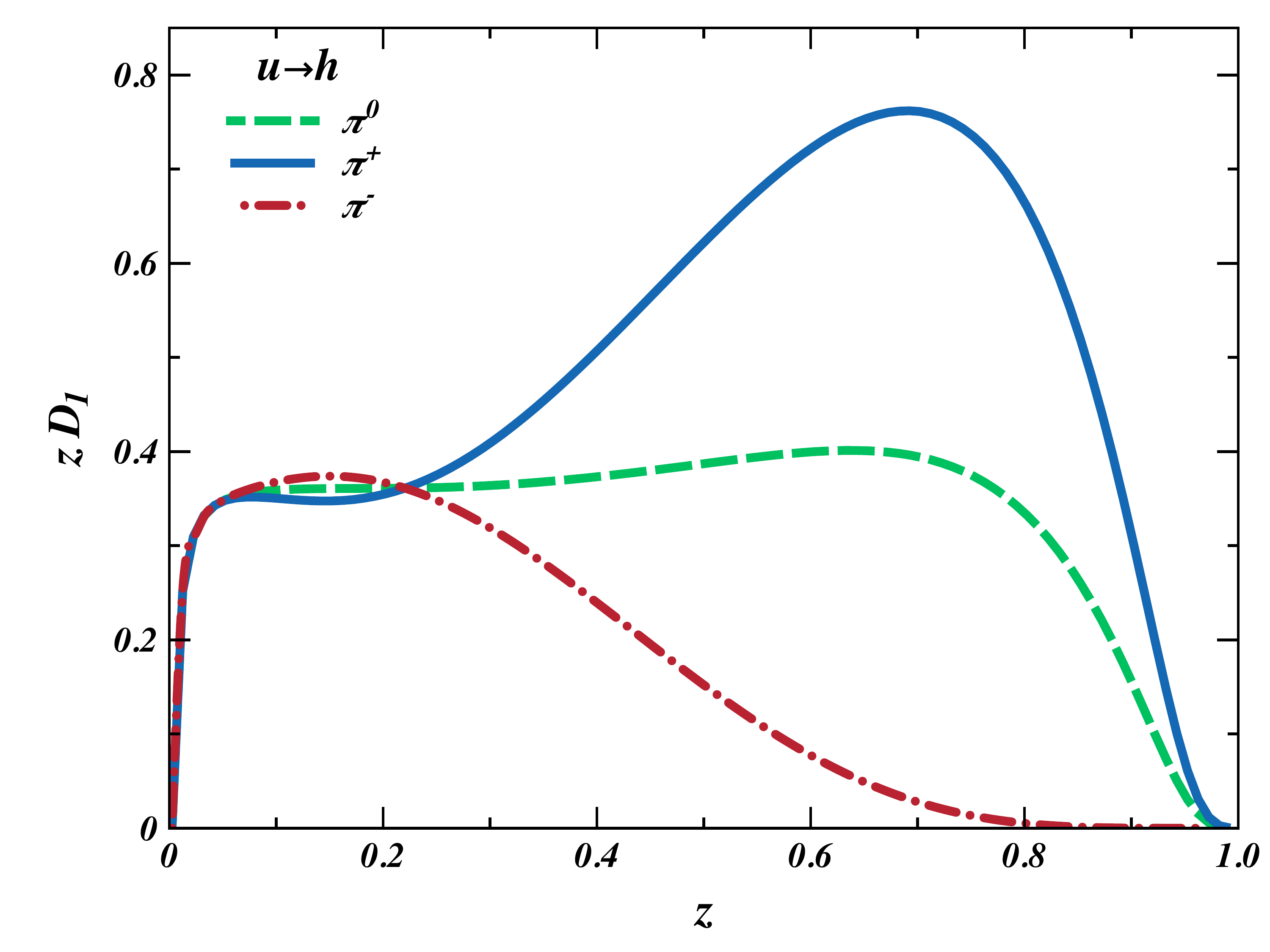}}
\hspace{0cm} 
\subfigure[] {
\includegraphics[width=0.9\columnwidth]{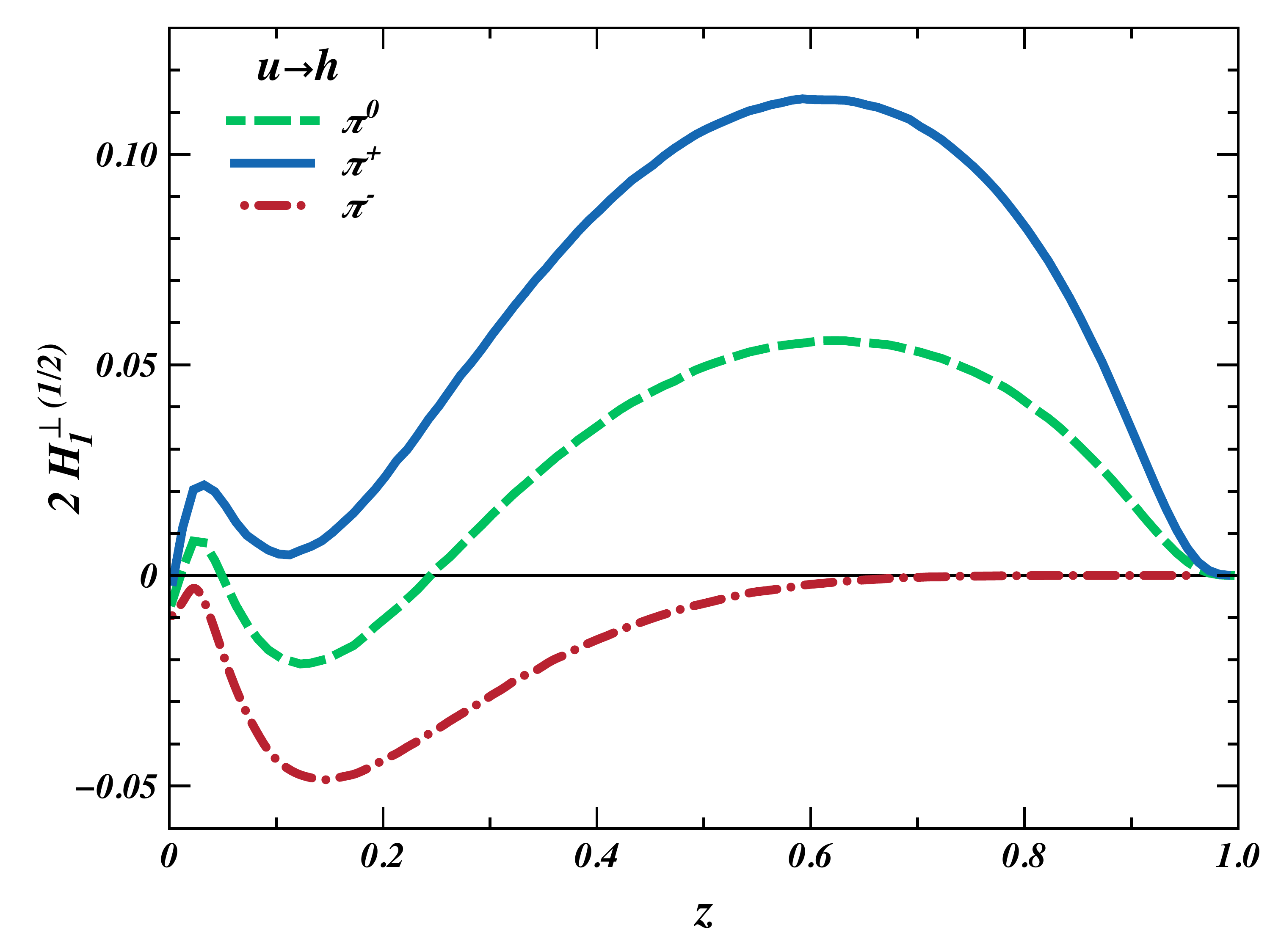}
}
\vspace{0.cm} 
\subfigure[] {
\includegraphics[width=0.9\columnwidth]{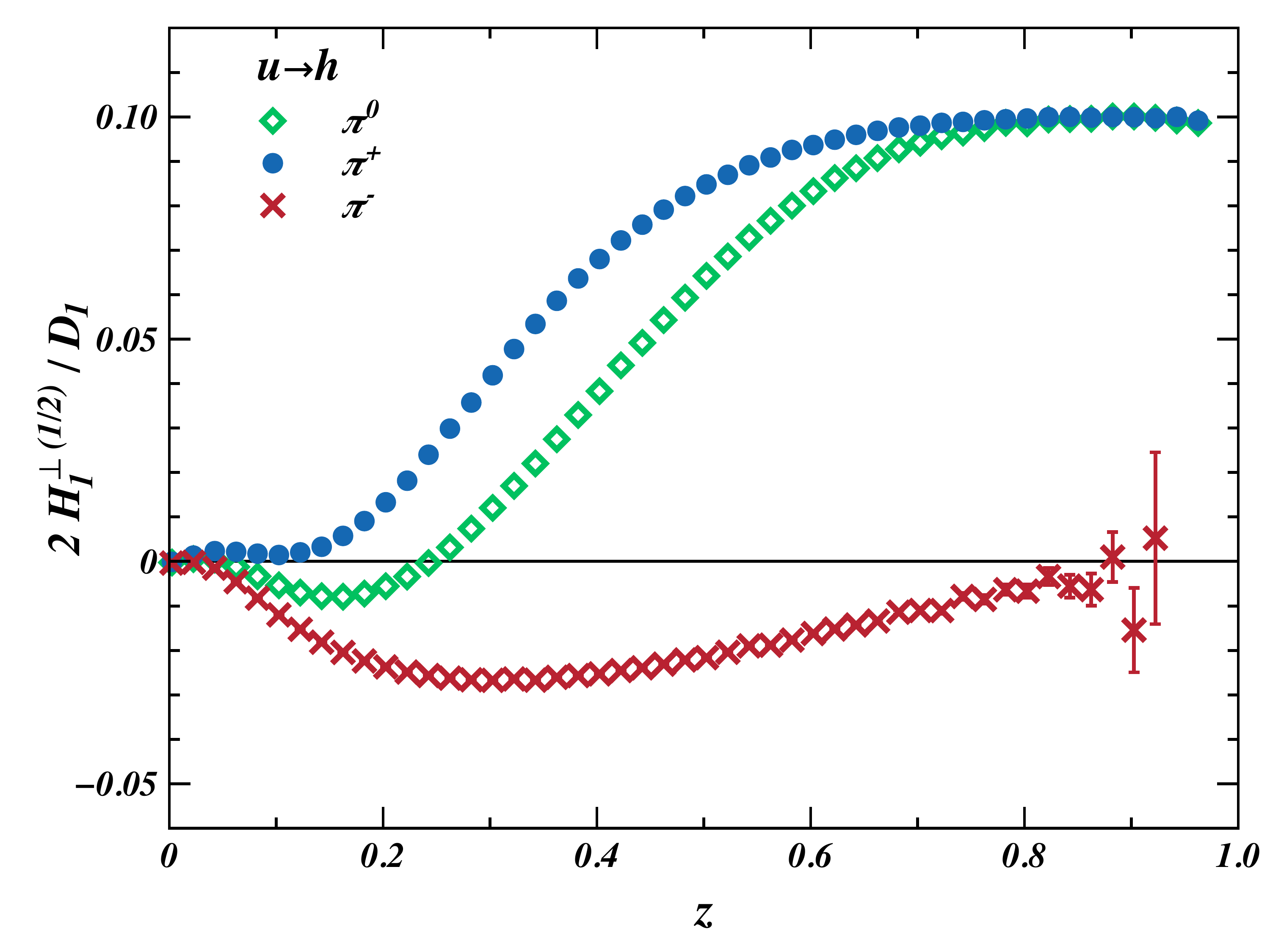}
}
\caption{Fitted values for unpolarized fragmentation function, $D_1$, (a) and twice the Collins function $1/2$ moment, $2 H_1^{\perp (1/2)}$, (b)  and their ratio (c) for $\pi^0$, $\pi^+$ and $\pi^-$ versus $z$, produced by $u$ quark from quark-jet framework using a toy model. The error bars show the uncertainties from the statistics and fits, and are only visible for the unfavored ratio at large values of $z$.}
\label{PLOT_D1_H12_TOY}
\end{figure}
 For the toy model, we perform the MC simulations with only light quarks in the quark-jet and pions in the final state, for simplicity. We perform several high-statistics simulations, where the number of produced hadrons in each decay chain, $N_{Links}$, is fixed to a particular value.  We use the relation~(\ref{EQ_MC_EXTRACT}) to extract the polarized number density $D_{h/q^{\uparrow}}(z,P_\perp^2,\varphi)$ from the numbers of produced hadrons in intervals of the variables $z$, $P_\perp^2$ and $\varphi$. For simplicity, in this section we will only consider the results for the $P_\perp^2$-integrated polarized quark fragmentation $D_{\pi^+/u^\uparrow}(z,\varphi)$ that can be obtained from $D_{h/q^{\uparrow}}(z,P_\perp^2,\varphi)$ using the relation in Eq.~(\ref{EQ_Nqh_Phi}). Next, to extract the unpolarized and the $1/2$ moment of the Collins functions  $D_1(z)$ and $H_1^{\perp (1/2)}(z)$,  we perform a minimum-$\chi^2$ fit  to $D_{h/q^{\uparrow}}(z,\varphi)$ using a form $F(c_0,c_1) \equiv c_0+c_1 \sin(\varphi)$ for fixed values of $z$. This fitting method allows us to better account for the statistical fluctuations in the MC results when extracting the Collins function. Notably, the fits describe the produced functions very well, yielding $\chi^2$ over the number of degrees of freedom always in the vicinity of $1$. A sample of such extraction for $u\to\pi^0$ fragmentation is shown in Fig.~\ref{PLOT_DPOL_TOY_PIP0}, for simulations with $N_{Links}=6$. The histograms (blue dots) show the polarized number density $D_{\pi^+/u^\uparrow}(z,\varphi)$ for two values of $z$ as a function of the azimuthal angle $\varphi$. The minimum-$\chi^2$ fits with a functional form $D_{h/{q^\uparrow}}= c_0+c_1 \sin (\varphi)$  (red lines) are also depicted. 

 Then, using $D_1(z)=2\pi c_0$ and  $-2 S_q H_1^{\perp (1/2)}(z)=2\pi c_1$ [see Eq.~(\ref{EQ_Nqh_Phi})], we plot the resulting fragmentation functions for $u$ quark and $N_{Links}=6$ as function of $z$ in Fig.~\ref{PLOT_D1_H12_TOY}, setting, for example, $S_q=-1$. The plots in Fig.~\ref{PLOT_D1_H12_TOY}(a) depict the extracted, integrated fragmentation functions, which are in agreement with the results of the direct MC calculations in our previous work~\cite{Matevosyan:2010hh,Matevosyan:2011vj, Matevosyan:2011ey}.
\begin{figure}[b]
\centering 
\subfigure[] {
\includegraphics[width=0.9\columnwidth]{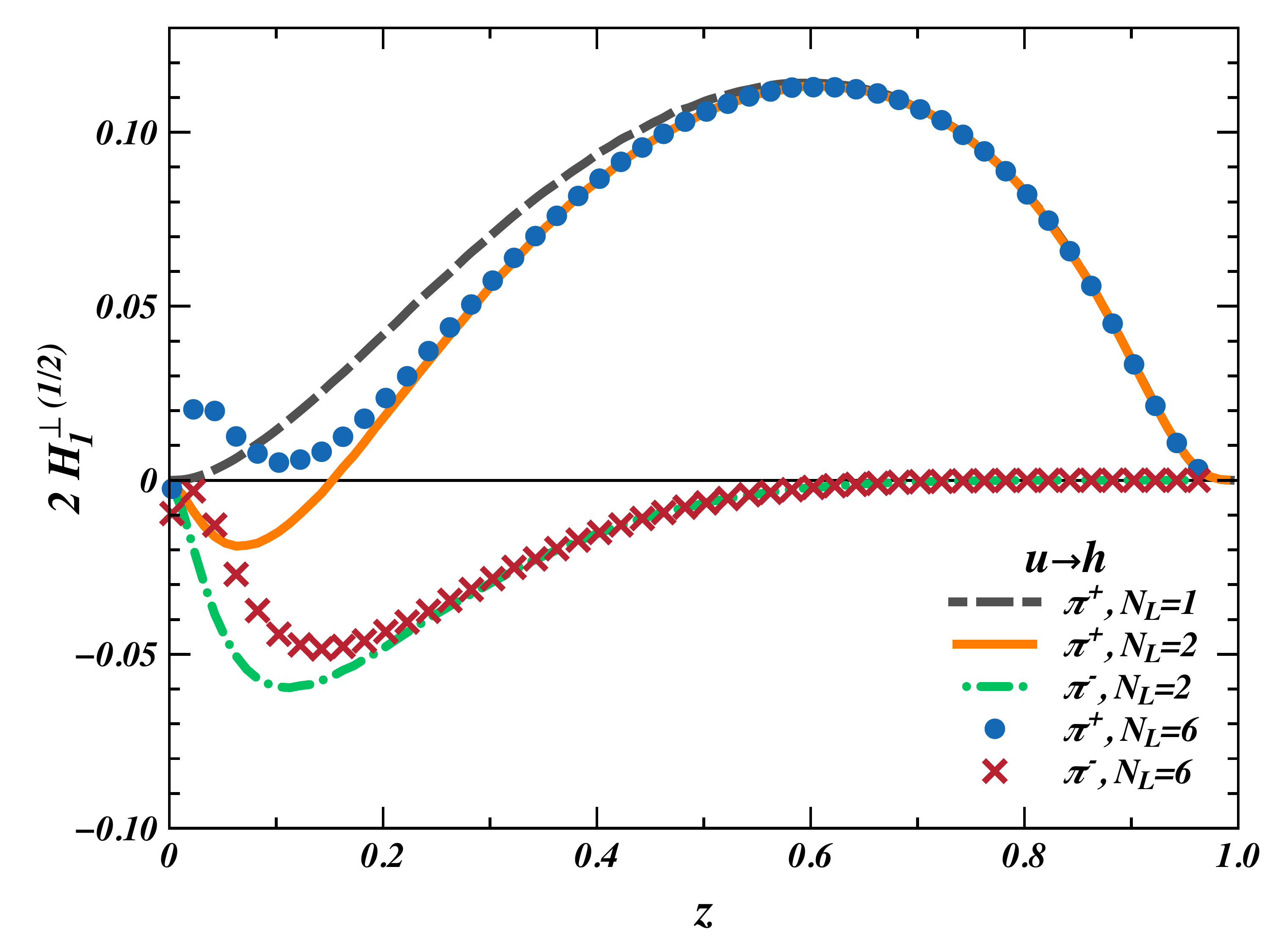}}
\hspace{0cm} 
\subfigure[] {
\includegraphics[width=0.9\columnwidth]{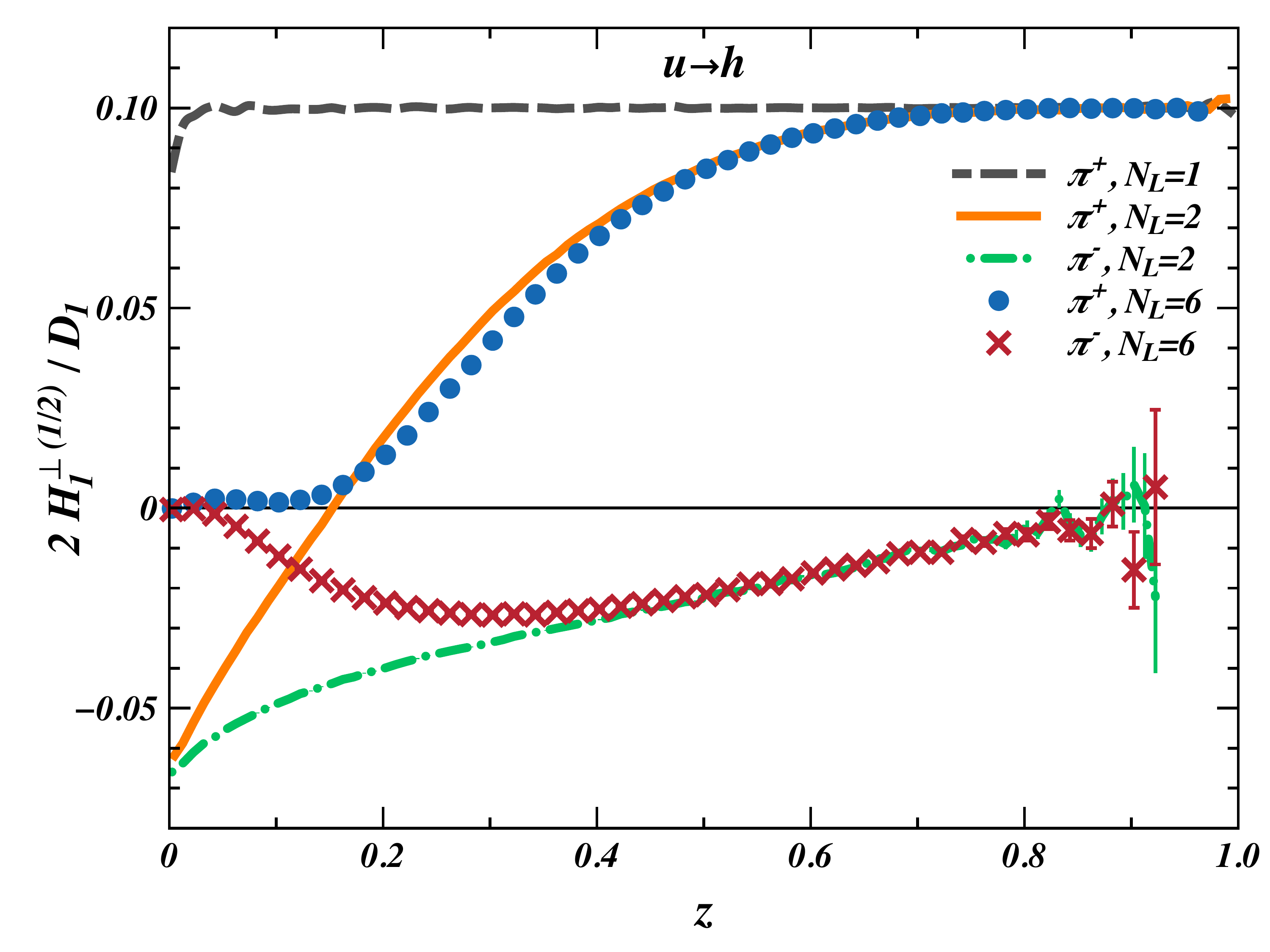}
}
\caption{Fitted values for $2 H_1^{\perp (1/2)}$  (a)  and the ratio $2H_1^{\perp (1/2)}/D_1$ (b) for $\pi^+$ and $\pi^-$ versus $z$, produced by $u$ quark from quark-jet framework using a toy model for increasing values of $N_L\equiv N_{Links}$.}
\label{PLOT_H12_BUILD_TOY}
\end{figure}
   
  The results for the moment of the Collins function, $2 H_1^{\perp (1/2)}$, depicted in Fig.~\ref{PLOT_D1_H12_TOY}(b), exhibit very interesting features: both functions for $u\to\pi^+$ and $u\to\pi^0$, though peak at values of $z\sim0.6$, decrease and $u\to\pi^0$ changes the sign for small values of $z$.  The unfavored function $u\to\pi^-$, generated solely by multihadron emission, has opposite sign and peak value at small $z$ of comparable size to that of the favored ones at large $z$. A similar picture is found for the ratio $2H_1^{\perp (1/2)}/D_1$, depicted in Fig.~\ref{PLOT_D1_H12_TOY}(c), where notably the results for $\pi^0$ are below those for $\pi^+$, while these ratios for the corresponding elementary functions coincide.

 To better understand the results for the $u\to\pi^+$ and $u\to\pi^-$ Collins functions, in Fig.~\ref{PLOT_H12_BUILD_TOY} we depict the results for $N_{Links}$ equal to $1$, $2$ and $6$. The plots for $2H_1^{\perp (1/2)}$ in Fig.~\ref{PLOT_H12_BUILD_TOY}(a) show that the functions change very little for $N_{Links}>2$ in the high $z$ region. That is, the bulk of the model effects for $z\gtrsim 0.2$ can be described with just two hadron emissions. We verified numerically with lower statistics runs, that the further increase in $N_{Links}>6$ affects the functions only at extremely low values of $z$, below the numerical discretization size of $\Delta z=0.01$ used in this study. A na\"ive interpretation for the results is that the remnant quark has larger probability to have its spin antiparallel to that of the splitting quark, see Eq.~(\ref{EQ_SPIN_FLIP-NON}), and  typically has a small fraction of the initial light-cone momentum. Then in the emission step of the second hadron that affects the low $z$ region and generates the unfavored function, the angle $\varphi$ in the the Collins term in Eq.~(\ref{EQ_ELEM_Dqh_SIN}) in most cases will acquire an additional $\pi$ phase, yielding the results in Fig.~\ref{PLOT_H12_BUILD_TOY}. This picture repeats for further hadron emissions, creating "destructive interference" in the small-$z$ region, decreasing the Collins functions. The ratio  $2H_1^{\perp (1/2)}/D_1$, depicted in Fig.~\ref{PLOT_H12_BUILD_TOY}(b), also exhibits the contrast in the low-$z$ behavior of the unpolarized and Collins functions: while $D_1(z)$ grows roughly as $1/z$, the Collins function oscillates to $0$. It is worth mentioning that, for $N_{Links}=1$, the fitted value of the ratio in $u\to\pi^+$ equals to the one set in the toy model as input to the MC.   
\section{Collins function For Pions and Kaons}
\label{SEC_COLLINS_RES}

\vspace{-0.2cm}
 In this section we present the results of the full MC simulations with light and strange fragmenting quarks as well as pions and kaons as the produced hadrons. The elementary splitting functions were taken from Eq.~(\ref{EQ_ELEM_Dqh_SIN}) and the simulations were done for values of $N_{Links}$ equal to $1$, $2$ and $6$. Again, it was checked with lower statistics runs, that the solutions are indistinguishable for $N_{Links}>6$ with the number of discretization points for $z$, $P_\perp^2$ and $\varphi$ used here.

\subsection{$P_\perp^2$ integrated results}
\label{SEC_SUB_FULL_RES_PT_INT}
 We first present the results for the $P_\perp^2$-integrated Collins function, skipping the results for the unpolarized functions, $D_1(z)$, as they have been studied in detail in our previous work. 
 
\begin{figure}[t]
\centering 
\subfigure[] {
\includegraphics[width=0.9\columnwidth]{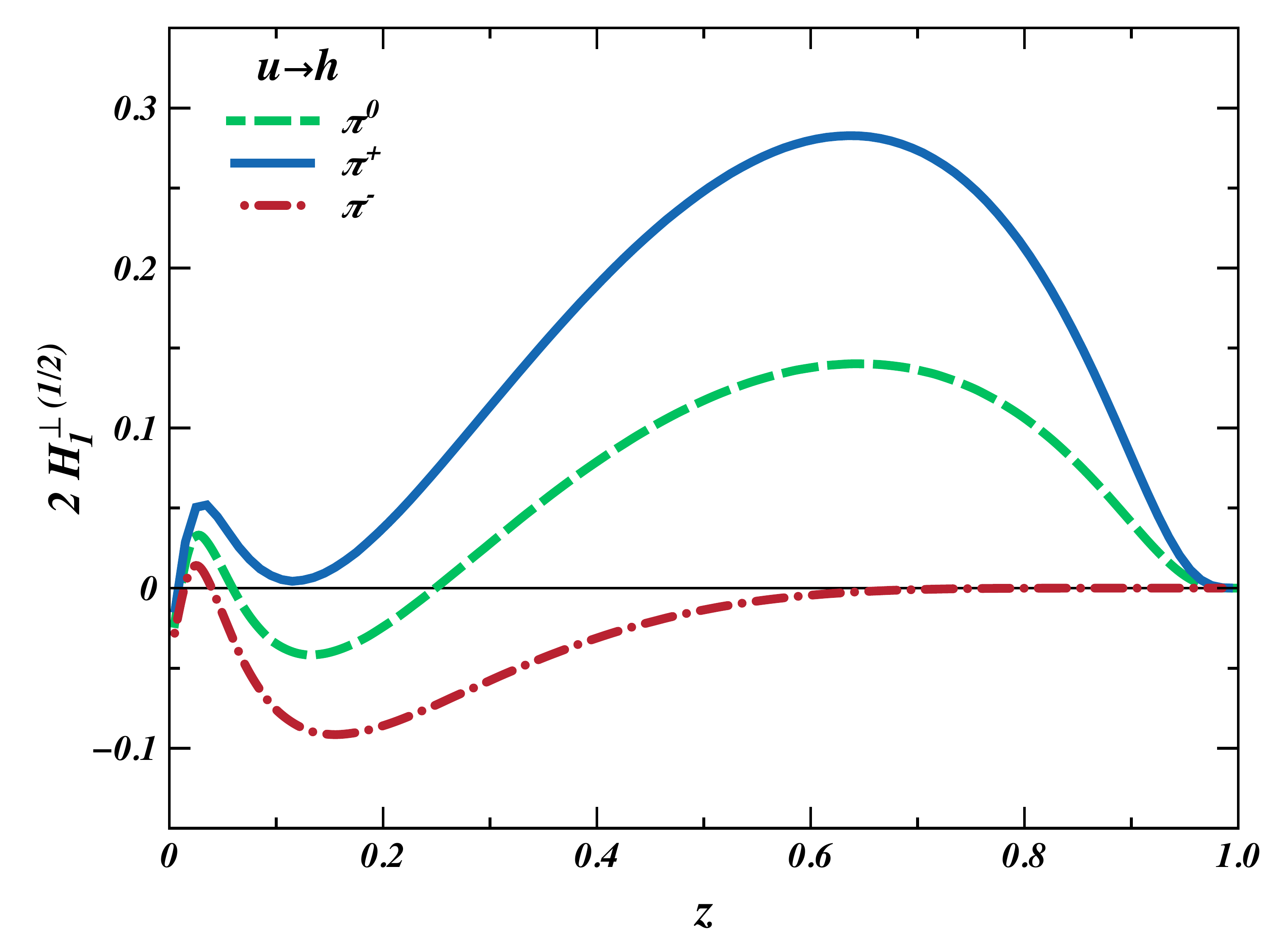}
}
\hspace{0cm} 
\subfigure[] {
\includegraphics[width=0.9\columnwidth]{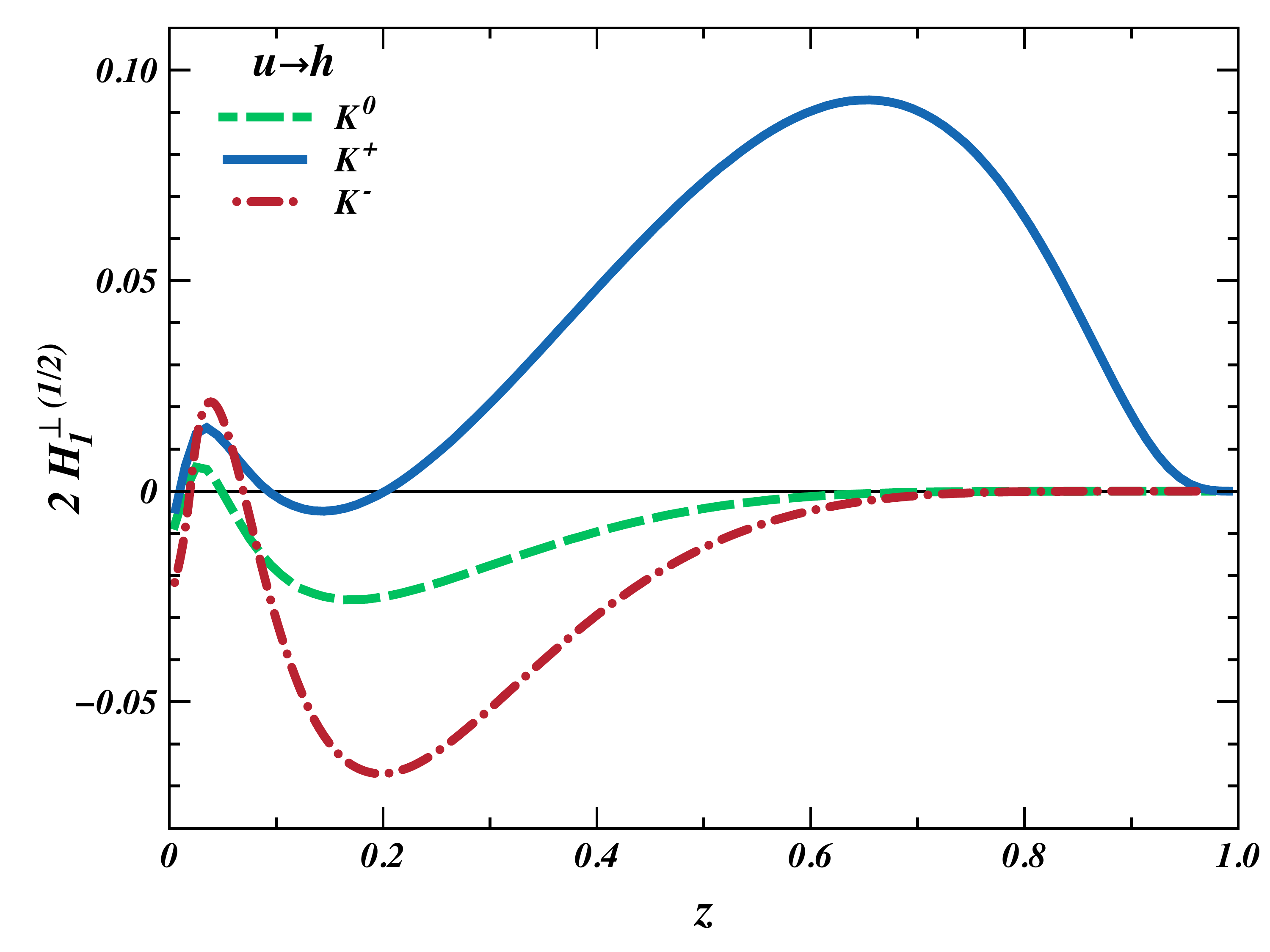}
}
\caption{Fitted values for $2H_1^{\perp (1/2)}$ for pions  (a)  and kaons  (b), produced by $u$ quark for $N_{Links}=6$.}
\label{PLOT_H12_U}
\end{figure}

\begin{figure}[t]
\centering 
\subfigure[] {
\includegraphics[width=0.9\columnwidth]{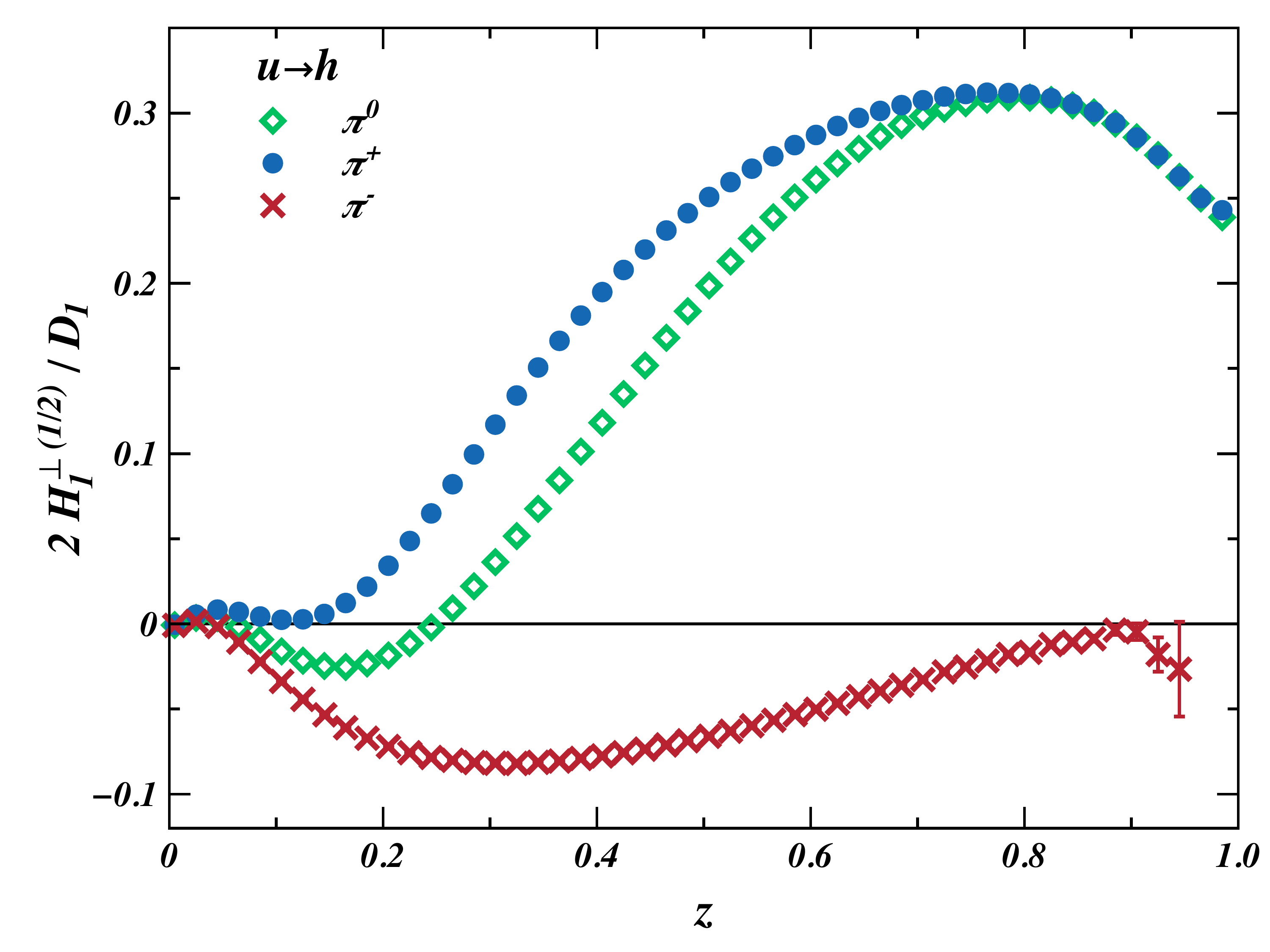}
}
\hspace{0cm} 
\subfigure[] {
\includegraphics[width=0.9\columnwidth]{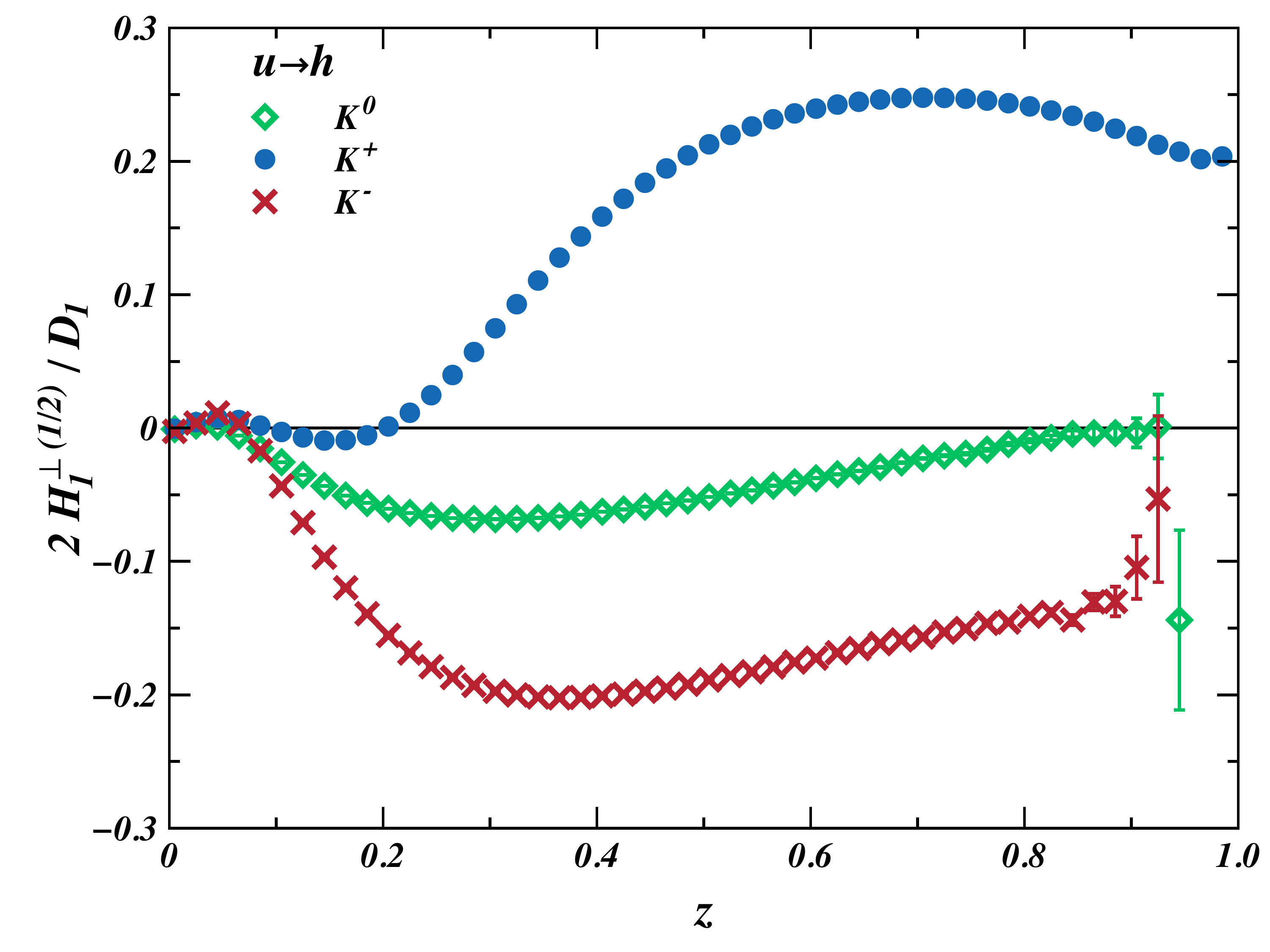}
}
\caption{The ratio of the fitted values, $2H_1^{\perp (1/2)}/D_1$, for pions  (a)  and kaons  (b), produced by $u$ quark for $N_{Links}=6$.}
\label{PLOT_RAT_U}
\end{figure}

 The results for $2H_1^{\perp (1/2)}$ and the ratio $2H_1^{\perp (1/2)}/D_1$ for the hadrons produced by a $u$ quark are shown in Figs.~\ref{PLOT_H12_U} and \ref{PLOT_RAT_U}, respectively. The fitted values for $2H_1^{\perp (1/2)}$ for pions show similar features to those from the toy model: the favored functions, positive and peaking at $z\sim 0.65$, decrease and some change sign at lower values of $z$, oscillating around zero as $z\to0$. The unfavored function, $u\to\pi^-$, is mostly negative and peaking at $z\sim 0.2$, with a peak value  about one-third of that for $u\to\pi^+$. The results for kaons are similar to the those for pions, with the exception of $u\to K^+$. This is a favored fragmentation, but the peak value of the unfavored function $u\to K^-$ is only slightly less of that for the favored fragmentation. Also, since in our model charge and isospin symmetries are exact, the results for $u\to \bar{K}^0$ coincide with the ones for $u\to {K}^-$ and are omitted.

\begin{figure}[phtb]
\centering 
\subfigure[] {
\includegraphics[width=0.9\columnwidth]{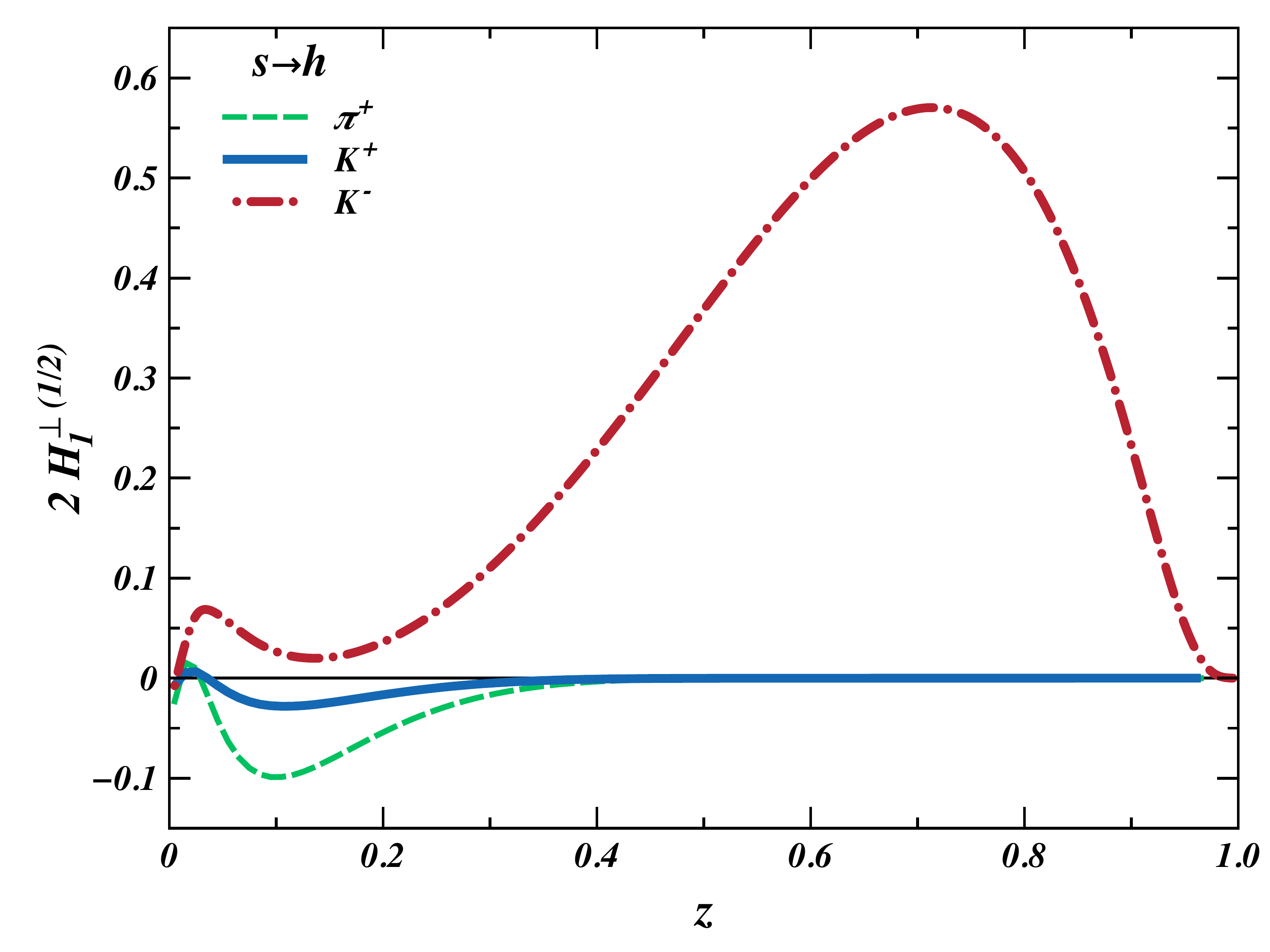}
}
\hspace{0cm} 
\subfigure[] {
\includegraphics[width=0.9\columnwidth]{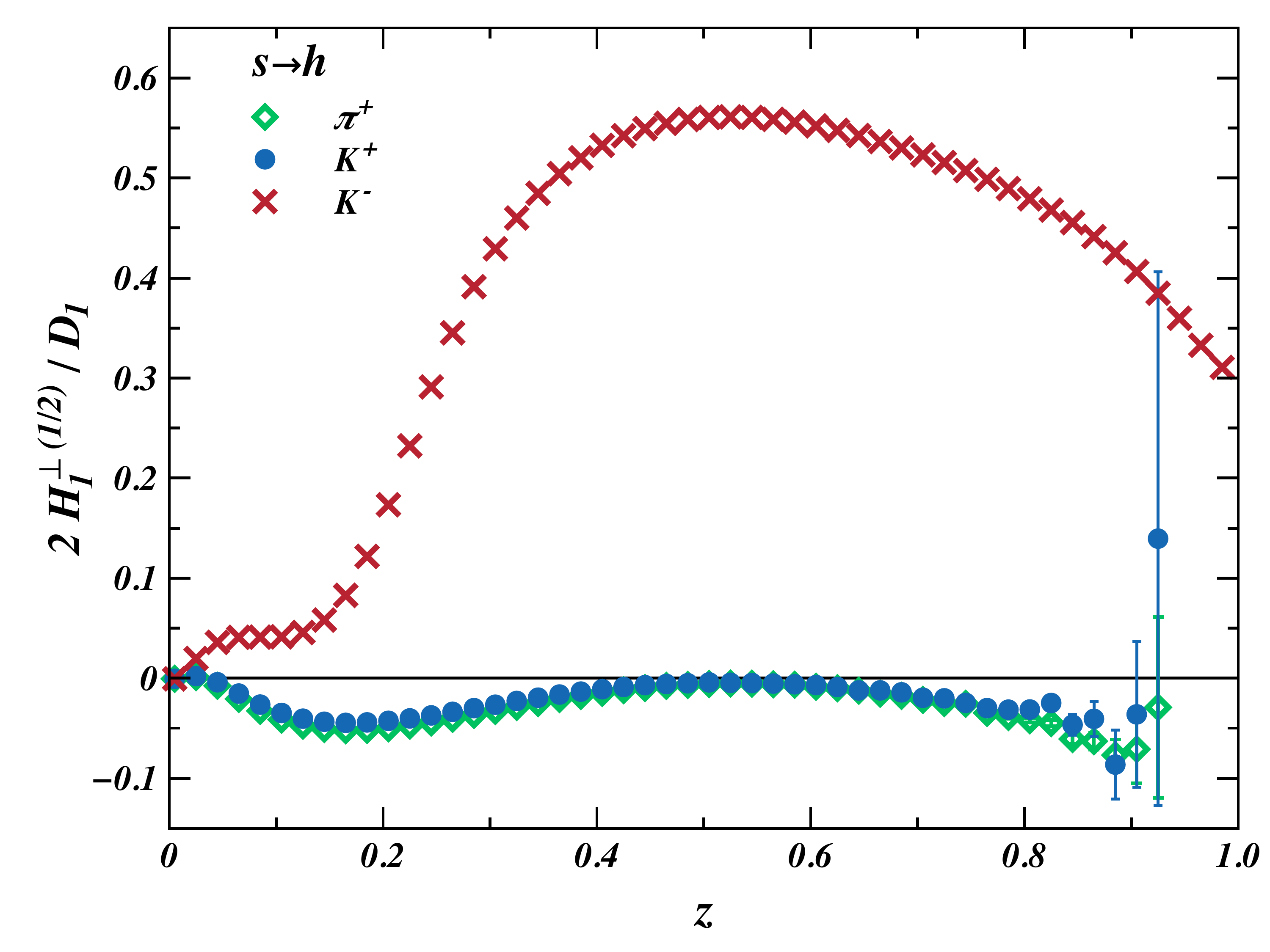}
}
\caption{Fitted values for $2H_1^{\perp (1/2)}$  (a)  and $2H_1^{\perp (1/2)}/D_1$  (b), for hadrons produced by $s$ quark for $N_{Links}=6$.}
\label{PLOT_H12_S}
\end{figure}

 Similar results for the $s$ quark are shown in Fig.~\ref{PLOT_H12_S}. The fitted values for $2H_1^{\perp (1/2)}$ for pions and unfavored kaon channels  peak at value of $z\simeq 0.1$ with values much smaller than those for the favored channels of $K^-$ and $\bar{K}^0$, which have a very broad peak and do not become negative at any value of $z$. 

\begin{figure}[phtb]
\centering 
\includegraphics[width=0.9\columnwidth]{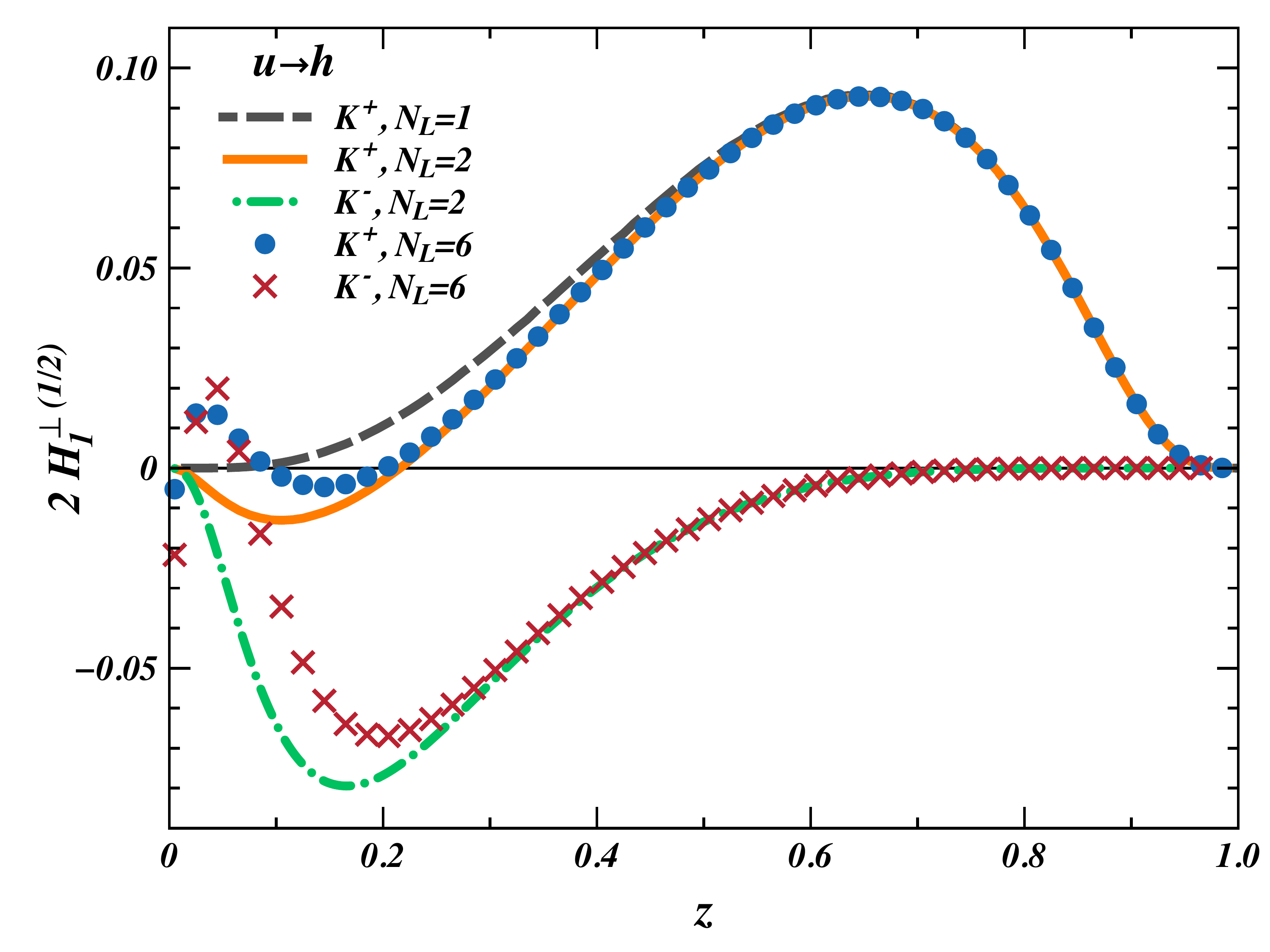}
\caption{Fitted values for $2H_1^{\perp (1/2)}$ for $K^+$ and $K^-$ versus $z$, produced by $u$ for increasing values of $N_L\equiv N_{Links}$.}
\label{PLOT_H12_U_K_NX}
\end{figure}

 The dependence on $N_{Links}$ can be seen in Fig.~\ref{PLOT_H12_U_K_NX}. Again, as for the toy model, we notice that the results change very little for $z>0.2$ for $N_{Links}\geqslant 2$. The increase in $N_{Links}$ affects the functions at decreasingly smaller values of $z$.

\vspace{-0.2cm}
\subsection{TMD of Collins function}
\label{SEC_SUB_TMD_COLLINS}

In this section we present the full results for the  TMD Collins function. As the $z$ dependence has been studied in detail in the previous sections, here we will only show the $P_\perp^2$ dependence of the Collins function for several fixed values of $z$. We noticed that the polar angle dependencies of the polarized number densities are not very accurately described by just the sine modulation term corresponding to the Collins term for some values of $z$ and $P_\perp^2$. In our calculations with $N_{Links}=6$, we found that a fourth order polynomial in $\sin(\varphi)$ 
\begin{align}
\label{EQ_COL_GEN}
D_{h/q^{\uparrow}} (z,P_\perp^2,\varphi) = \sum_{n=0}^{4} c_n(z, P_\perp^2) \sin^n \varphi
\end{align}
is sufficient to achieve fits for all slices with fixed $z$, $P_{\perp}^2$ with $\chi^2$ per degree of freedom in the vicinity of $1$. Here we identify $c_0$ with the unpolarized and $c_1$ with the Collins term of Eq.~(\ref{EQ_Dqh_SIN}). The higher powers of the $\sin(\varphi)$ in the polarized quark fragmentation functions are induced by the multiple hadron emissions.  A toy model analysis showed that there are two sources for these effects: the remnant quark transverse momentum distribution modulation from the recoil in the elementary hadron emissions,  and an additional modulation of the same quark distribution due to the $\varphi$ dependence of the quark spin flip probabilities. The detailed description and explanation of these additional terms are presented in our forthcoming publications~\cite{Matevosyan:2012ms, Matevosyan:2012ed}, where we show that these effects are a genuine feature of the quark-jet model and do not depend on the particular form of the elementary polarized fragmentation function.  
\begin{figure}[t]
\centering 
\subfigure[] {
\includegraphics[width=0.9\columnwidth]{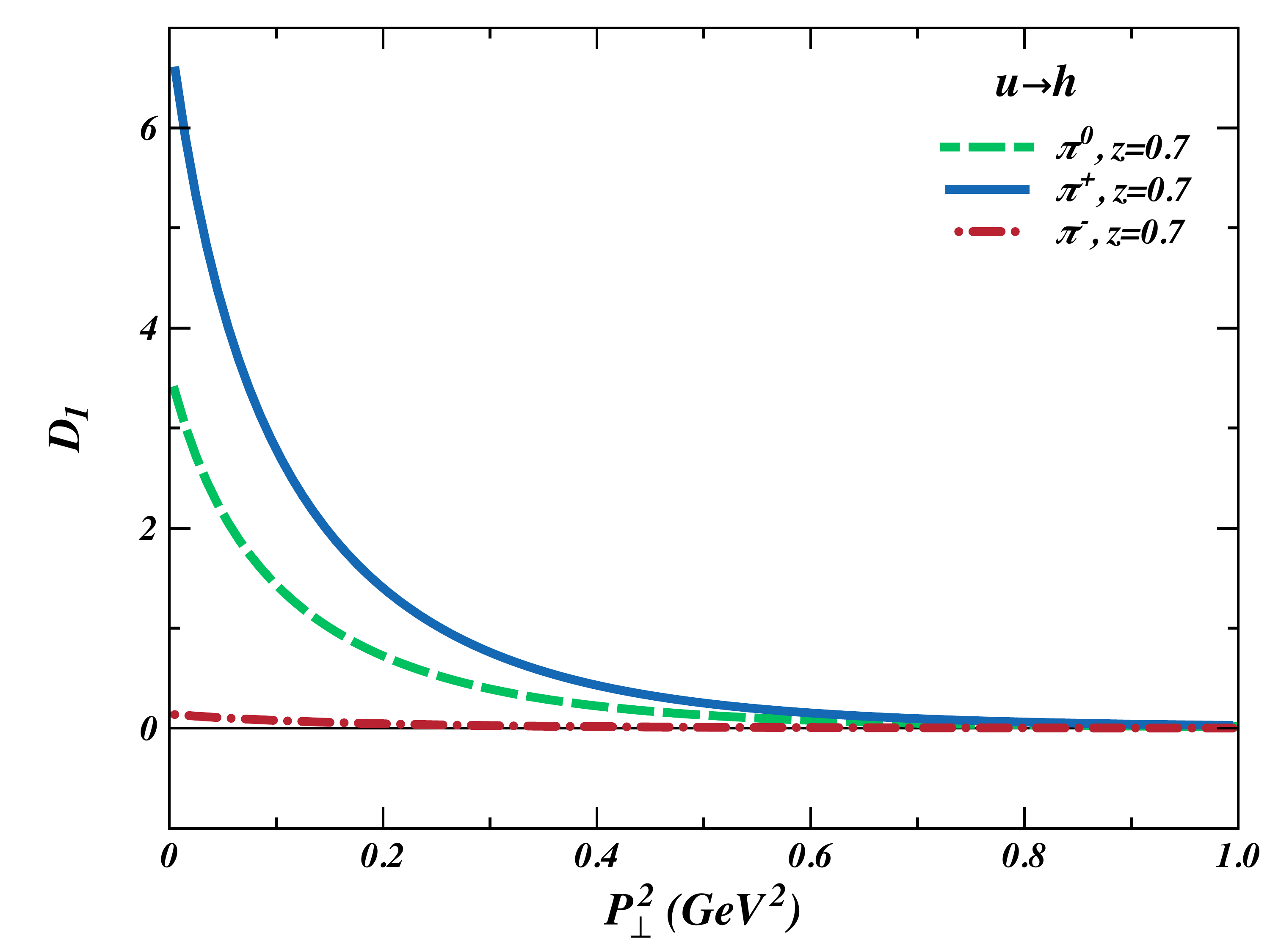}
}
\hspace{0cm} 
\subfigure[] {
\includegraphics[width=0.9\columnwidth]{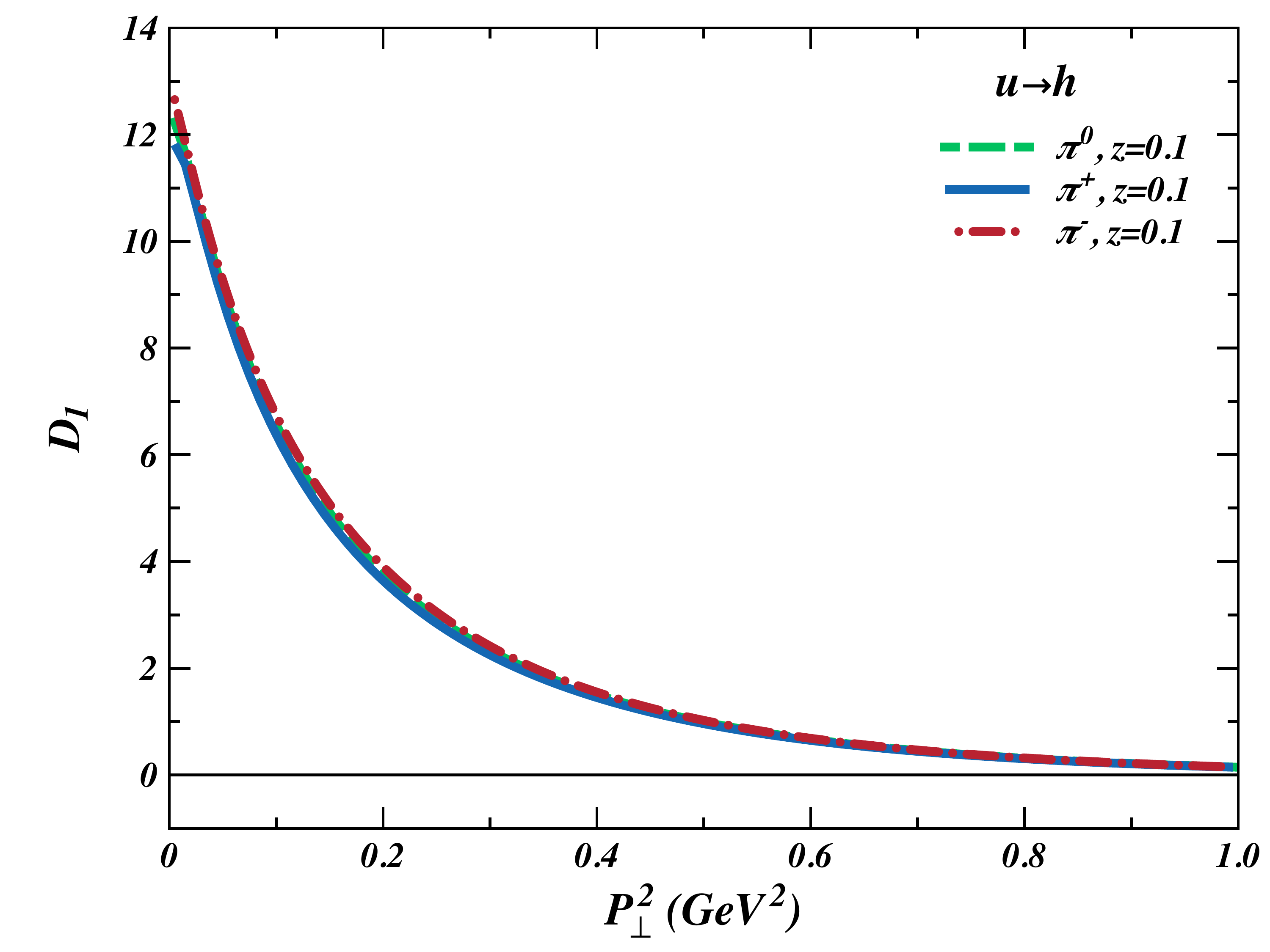}
}
\caption{The unpolarized fragmentation function vs $P_\perp^2$ for pions produced by a $u$ quark with $z=0.7$  (a)  and $z=0.1$ (b) for $N_{Links}=6$.}
\label{PLOT_TMD_D1_U_PI}
\end{figure}
   The plots in Figs.~\ref{PLOT_TMD_D1_U_PI},~\ref{PLOT_TMD_COL_U_PI} and~\ref{PLOT_TMD_RATCOL_U_PI} depict the results for unpolarized, Collins fragmentation functions and their ratios respectively for pions produced by an initial $u$ quark with two fixed values of $z$ equal to $0.7$ and $0.1$. For $z=0.7$, the results are mostly affected by a single hadron emission, thus the $P_\perp^2$ dependence of the Collins function is somewhat similar to one of the unpolarized fragmentations (which, in turn are reasonably well described via Gaussian function for small values of $P_\perp^2$ at a fixed $z$, see \cite{Matevosyan:2011vj}), but peak at a small nonzero value of $P_\perp^2$. The results are quite different for $z=0.1$, where the unpolarized functions for all the pions are roughly equal, while the Collins function for all the pions have similar shapes but differ in magnitude, decreasing to a negative value and peaking at $P_\perp^2 \sim 0.1 \mathrm{GeV^2}$, then start to increase and become positive for small values of $P_\perp^2$. Thus, it follows from our model that while the transverse momentum dependence of the unpolarized fragmentation functions can be to a good approximation described by a Gaussian function with $z$-dependent width, the Collins functions have much different shapes.   

\begin{figure}[t]
\centering 
\subfigure[] {
\includegraphics[width=0.9\columnwidth]{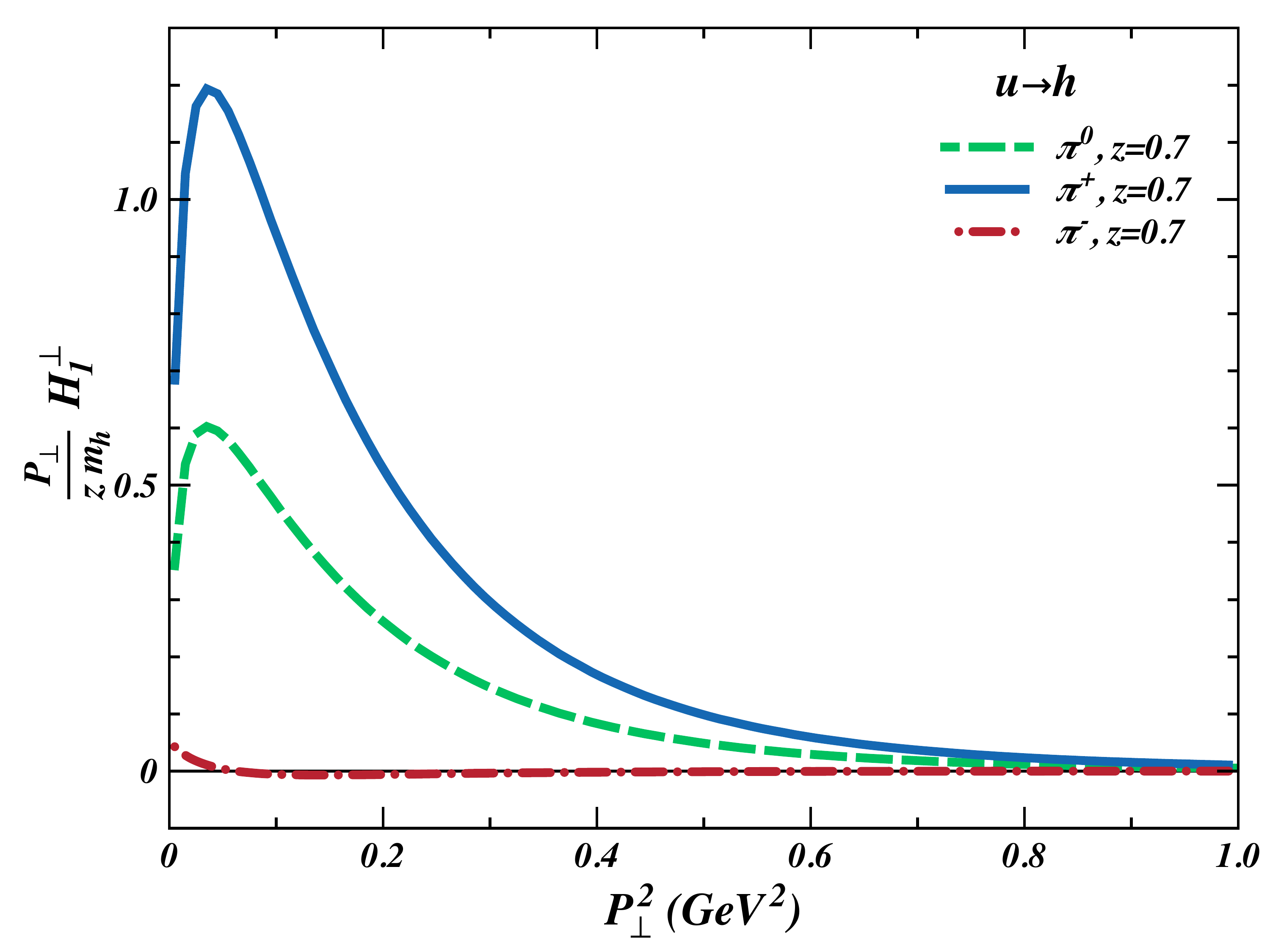}
}
\hspace{0cm} 
\subfigure[] {
\includegraphics[width=0.9\columnwidth]{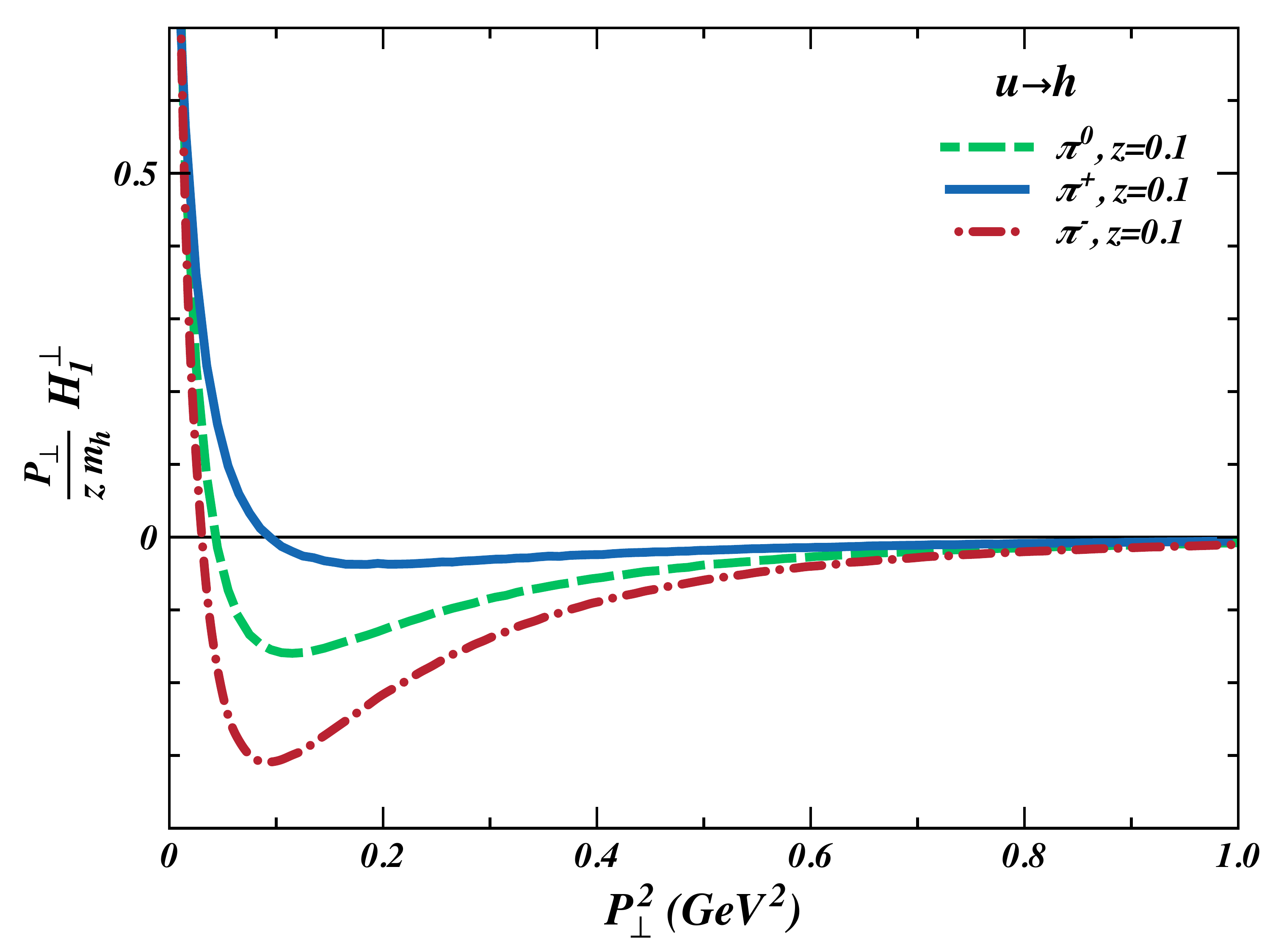}
}
\caption{The Collins fragmentation function vs $P_\perp^2$ for pions produced by a $u$ quark with $z=0.7$  (a)  and $z=0.1$ (b) for $N_{Links}=6$.}
\label{PLOT_TMD_COL_U_PI}
\end{figure}
\begin{figure}[phtb]
\centering 
\subfigure[] {
\includegraphics[width=0.9\columnwidth]{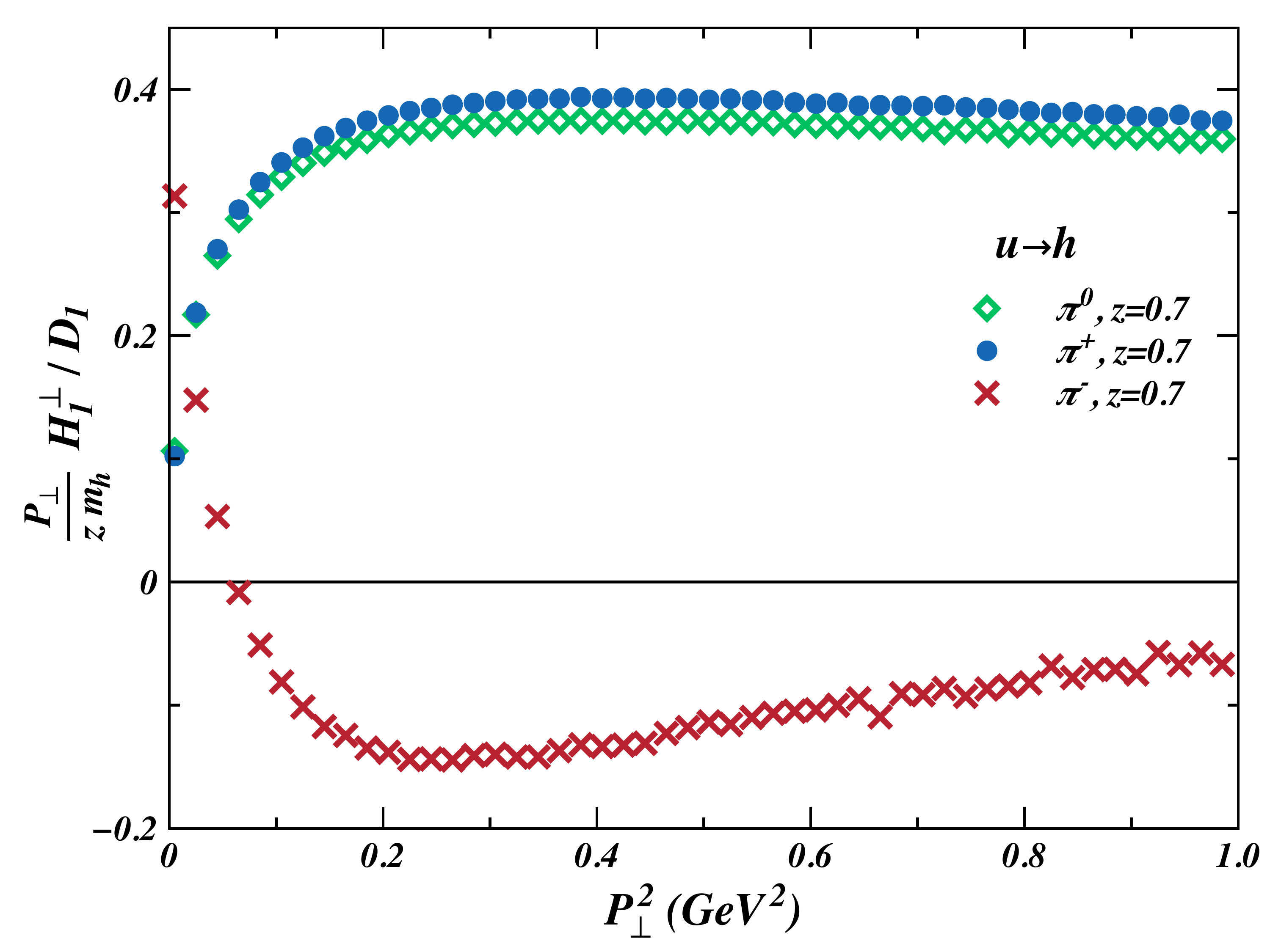}
}
\hspace{0cm} 
\subfigure[] {
\includegraphics[width=0.9\columnwidth]{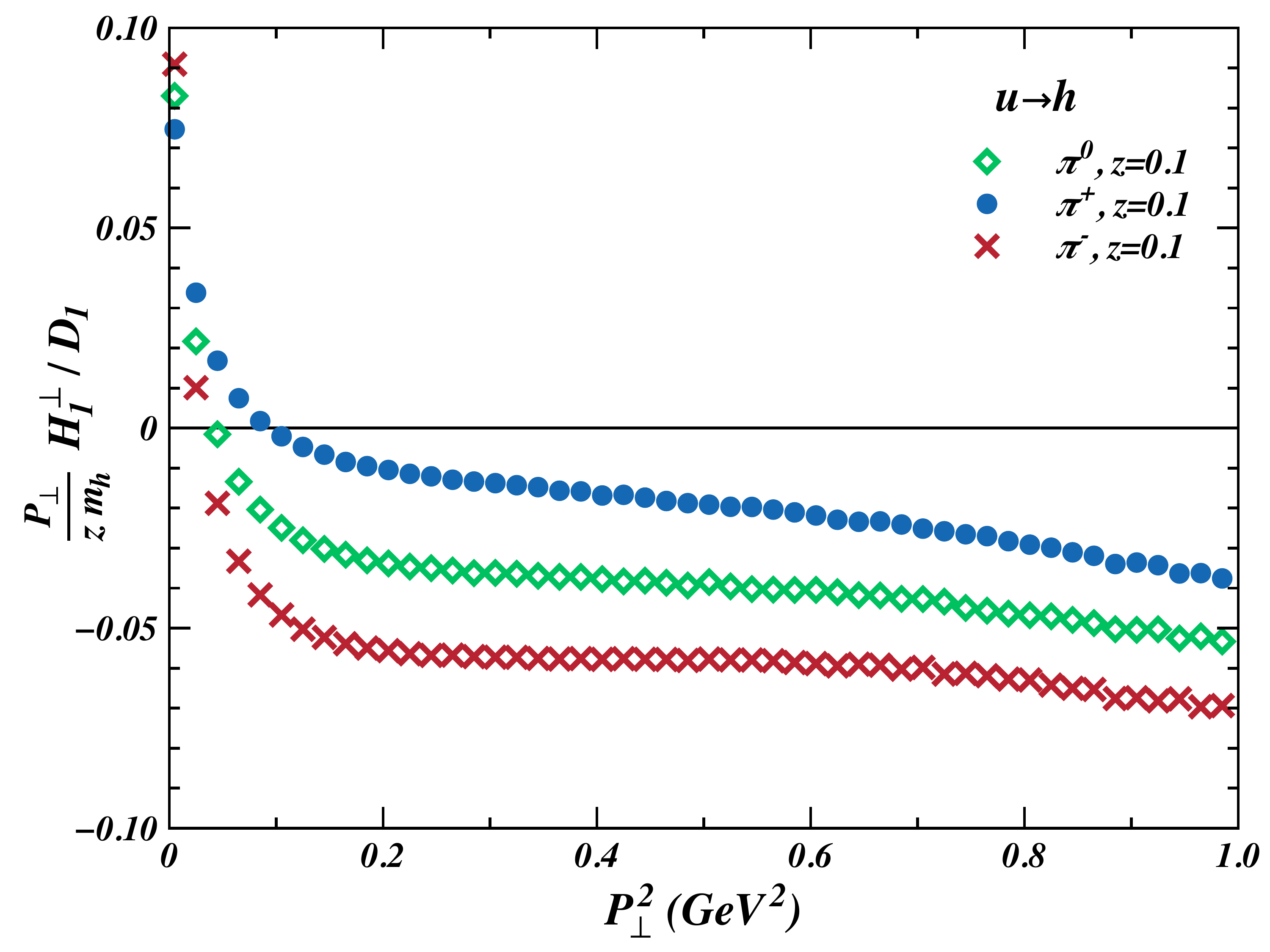}
}
\caption{The ratio of Collins fragmentation function to unpolarized fragmentation function vs $P_\perp^2$ for pions produced by a $u$ quark with $z=0.7$  (a)  and $z=0.1$ (b) for $N_{Links}=6$.}
\label{PLOT_TMD_RATCOL_U_PI}
\end{figure}

\begin{figure}[phtb]
\centering 
\subfigure[] {
\includegraphics[width=0.9\columnwidth]{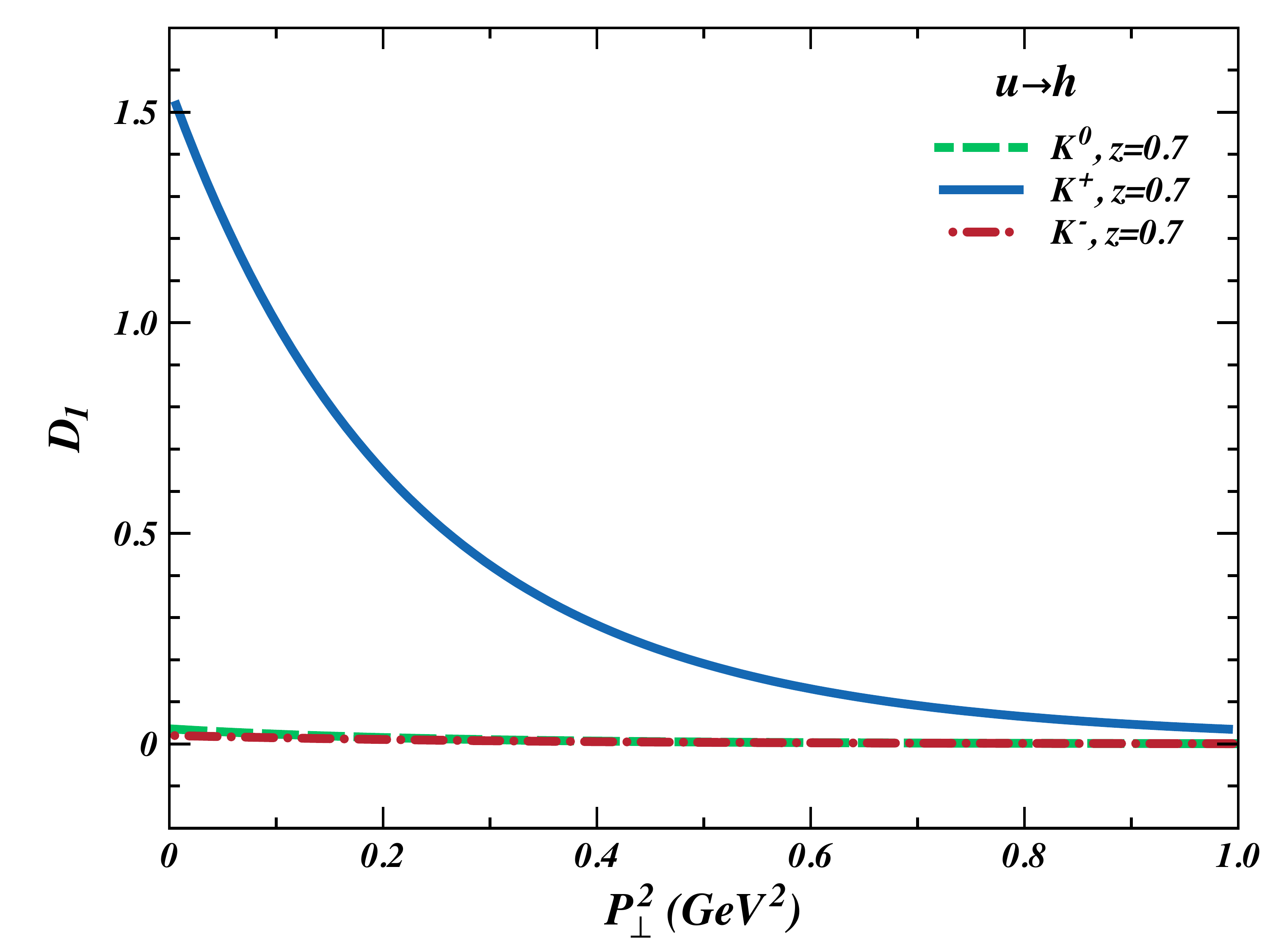}
}
\hspace{0cm} 
\subfigure[] {
\includegraphics[width=0.9\columnwidth]{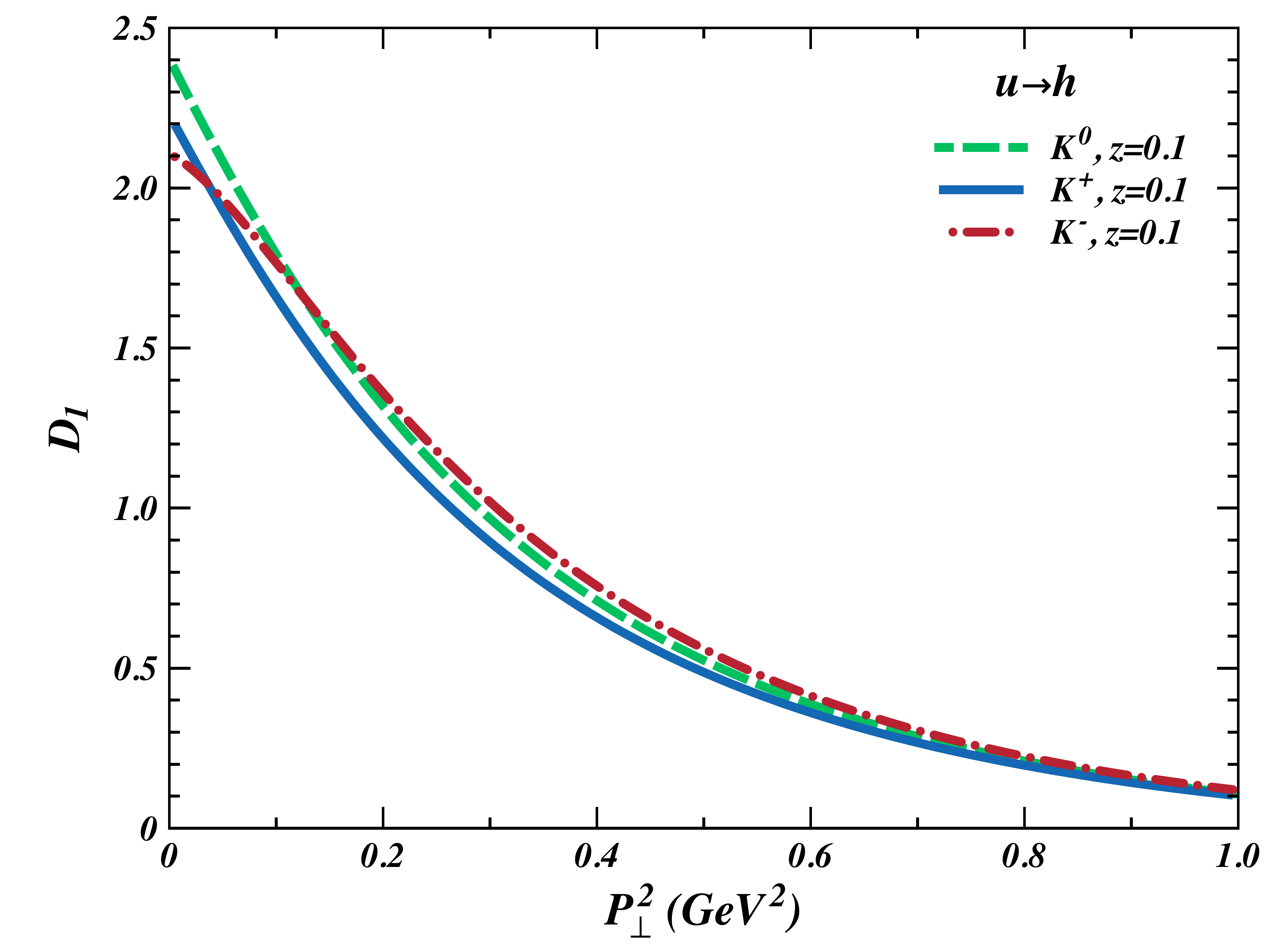}
}
\caption{The unpolarized fragmentation function vs $P_\perp^2$ for kaons produced by a $u$ quark with $z=0.7$  (a)  and $z=0.1$ (b) for $N_{Links}=6$.}
\label{PLOT_TMD_D1_U_K}
\end{figure}

\begin{figure}[tb]
\centering 
\subfigure[] {
\includegraphics[width=0.9\columnwidth]{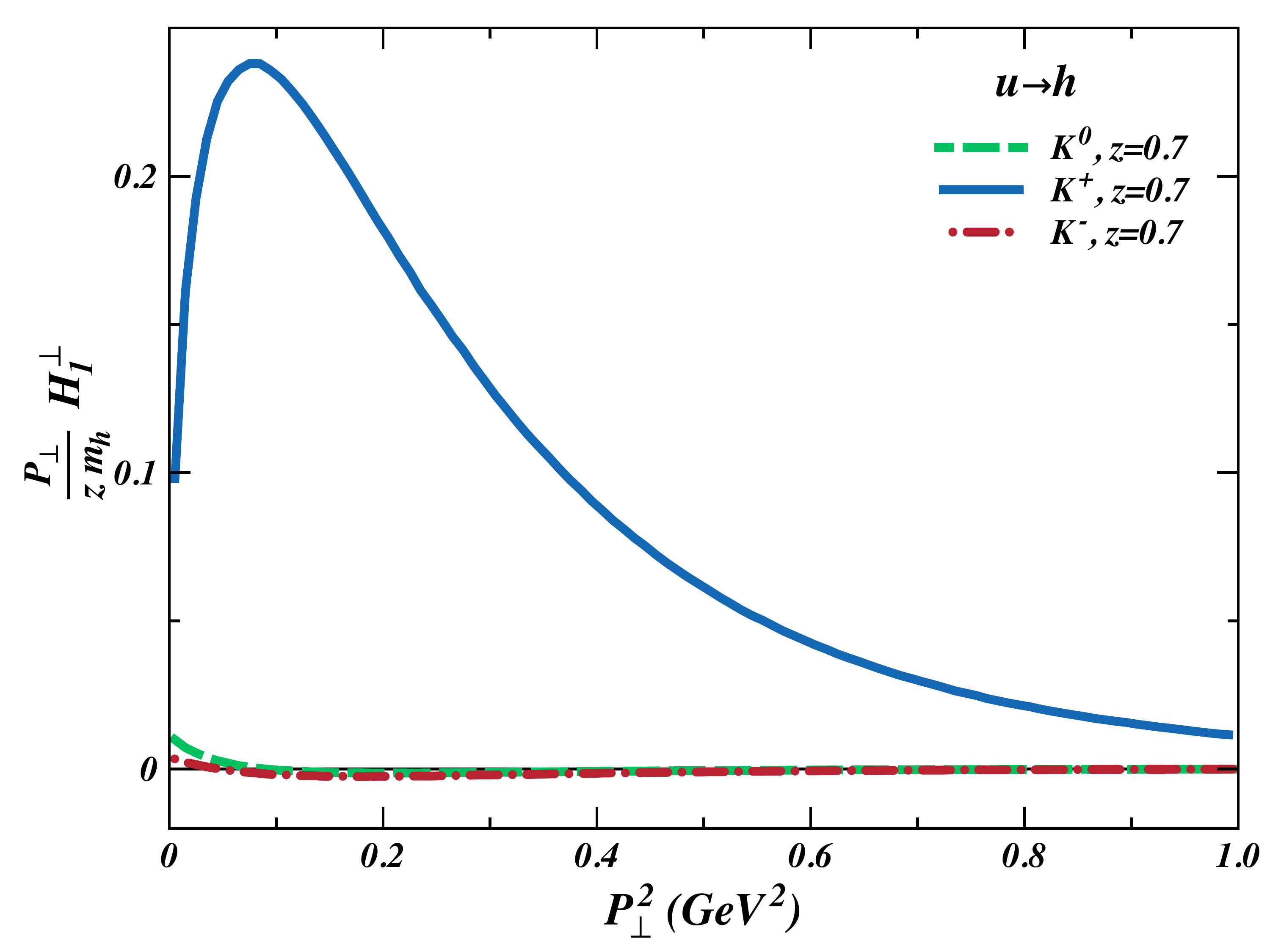}
}
\hspace{0cm} 
\subfigure[] {
\includegraphics[width=0.9\columnwidth]{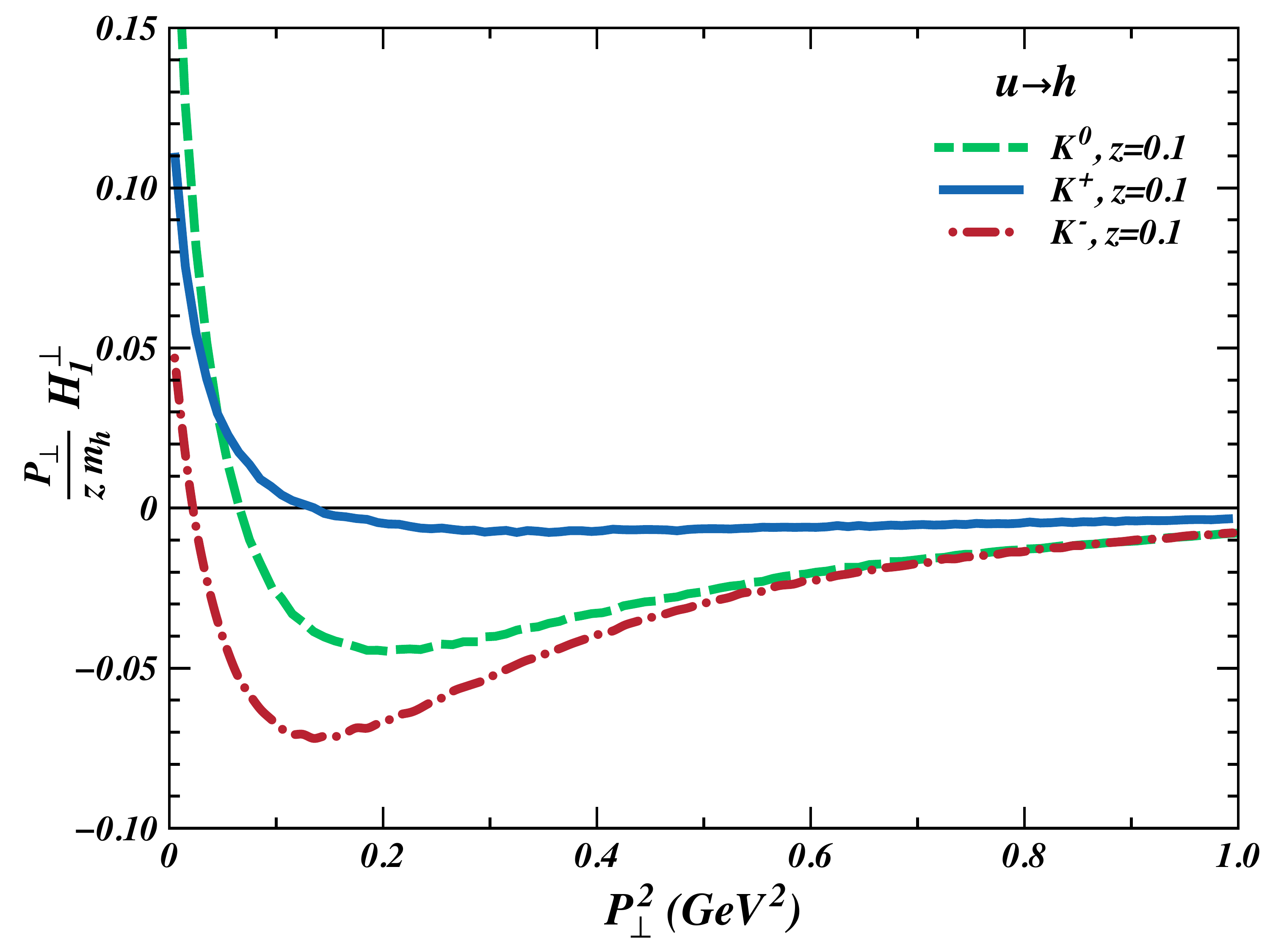}
}
\caption{The Collins fragmentation function vs $P_\perp^2$ for kaons produced by a $u$ quark with $z=0.7$  (a)  and $z=0.1$ (b) for $N_{Links}=6$.}
\label{PLOT_TMD_COL_U_K}
\end{figure}

\begin{figure}[t]
\centering 
\subfigure[] {
\includegraphics[width=0.9\columnwidth]{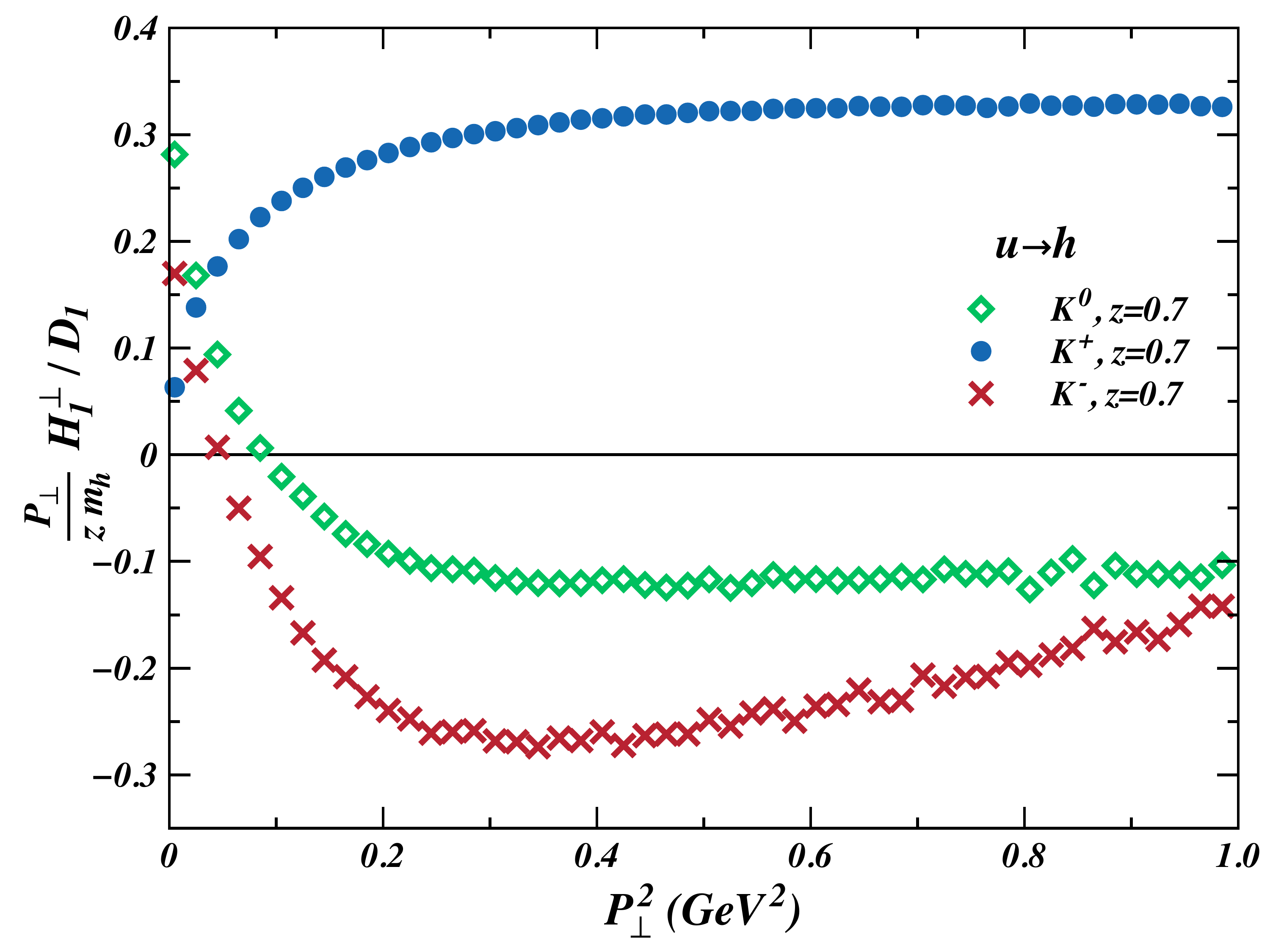}
}
\hspace{0cm} 
\subfigure[] {
\includegraphics[width=0.9\columnwidth]{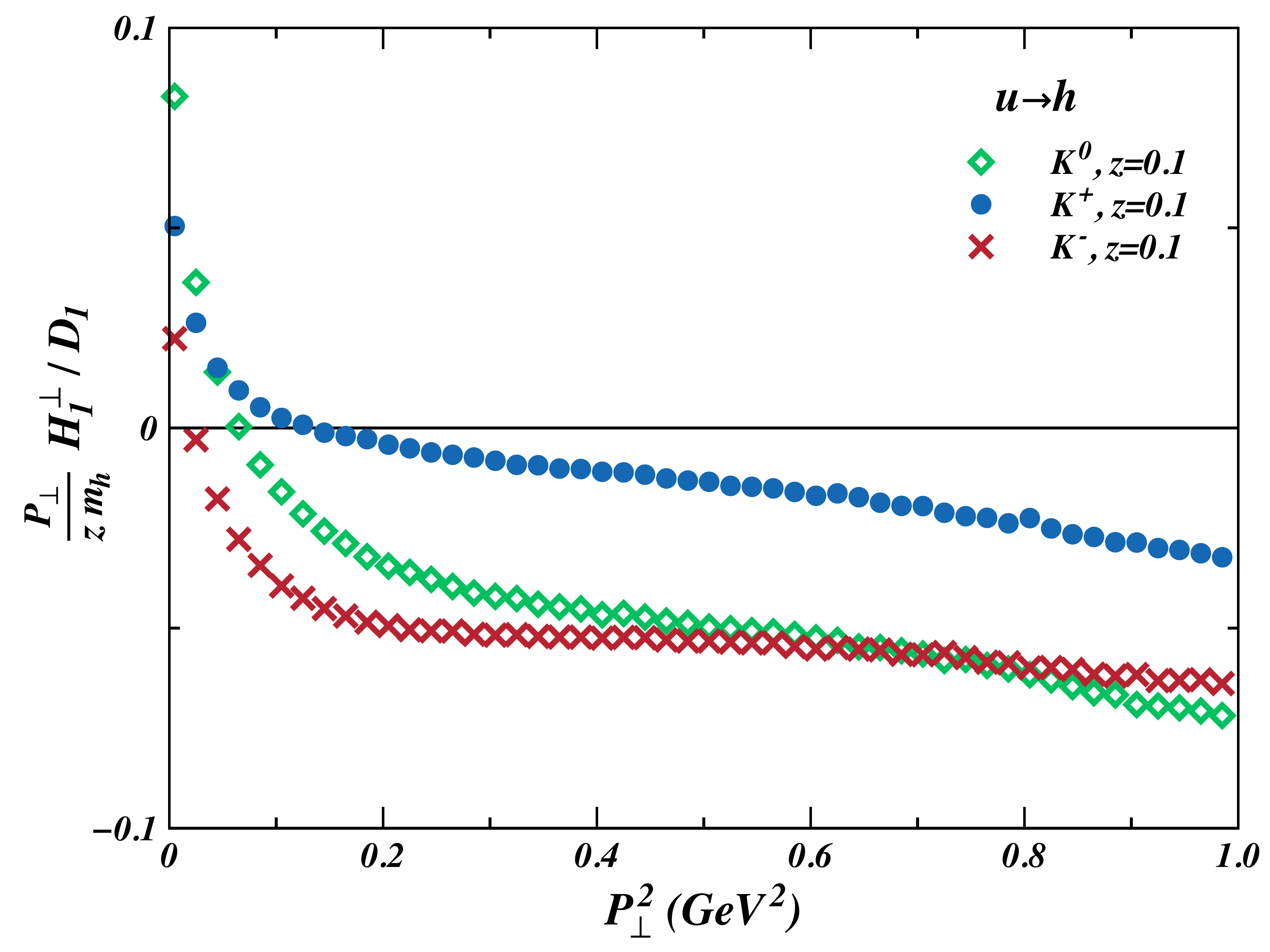}
}
\caption{The ratio of Collins fragmentation function to unpolarized fragmentation function vs $P_\perp^2$ for kaons produced by a $u$ quark with $z=0.7$  (a)  and $z=0.1$ (b) for $N_{Links}=6$.}
\label{PLOT_TMD_RATCOL_U_K}
\end{figure}

The plots in Figs.~\ref{PLOT_TMD_D1_U_K},~\ref{PLOT_TMD_COL_U_K}  and ~\ref{PLOT_TMD_RATCOL_U_K} depict the analogous results,  but for kaons produced by an initial $u$ quark. The results are similar to those for the pions, except for the magnitudes of the unpolarized and Collins functions being smaller.

\begin{figure}[t]
\centering 
\subfigure[] {
\includegraphics[width=0.9\columnwidth]{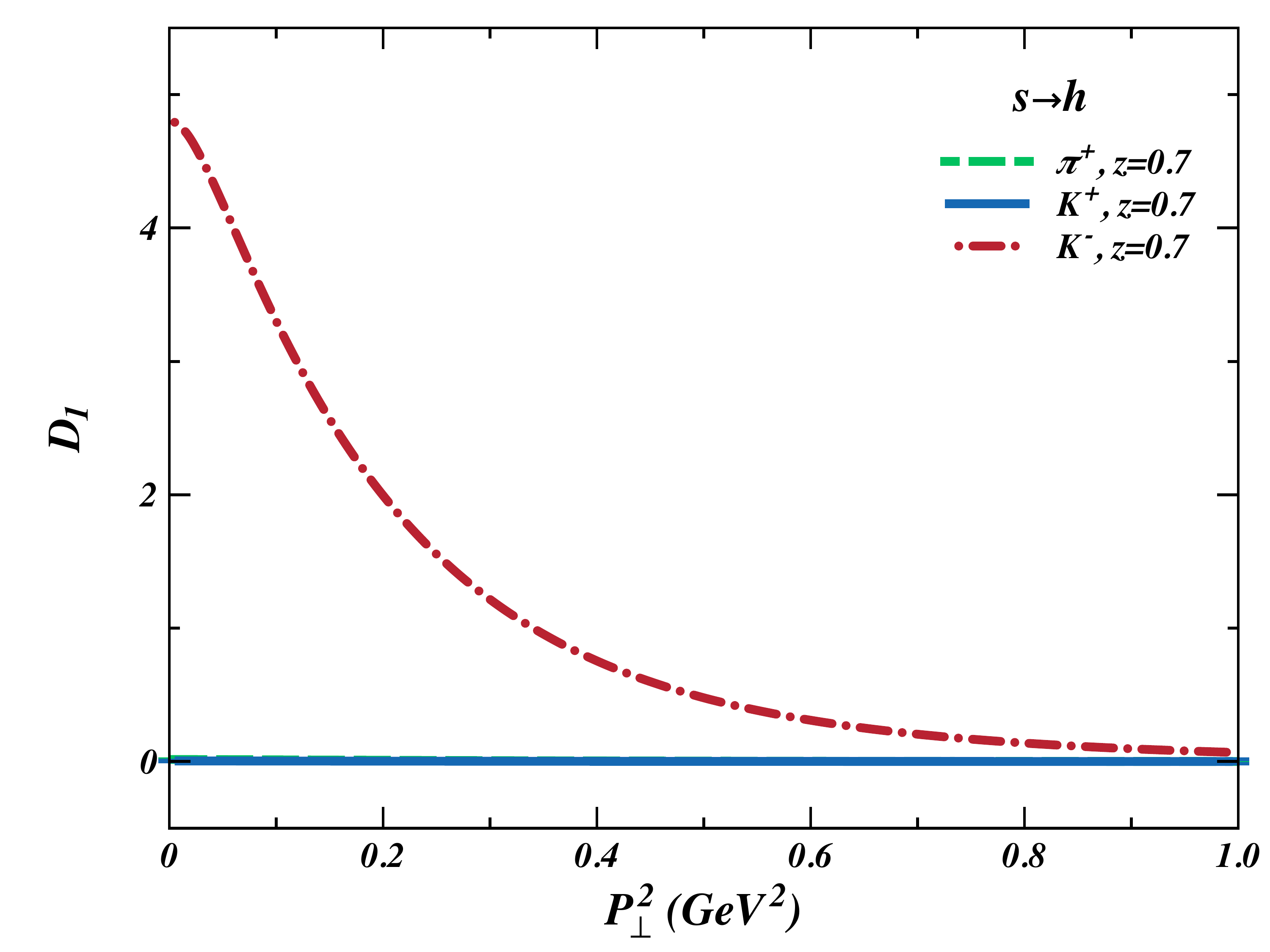}
}
\hspace{0cm} 
\subfigure[] {
\includegraphics[width=0.9\columnwidth]{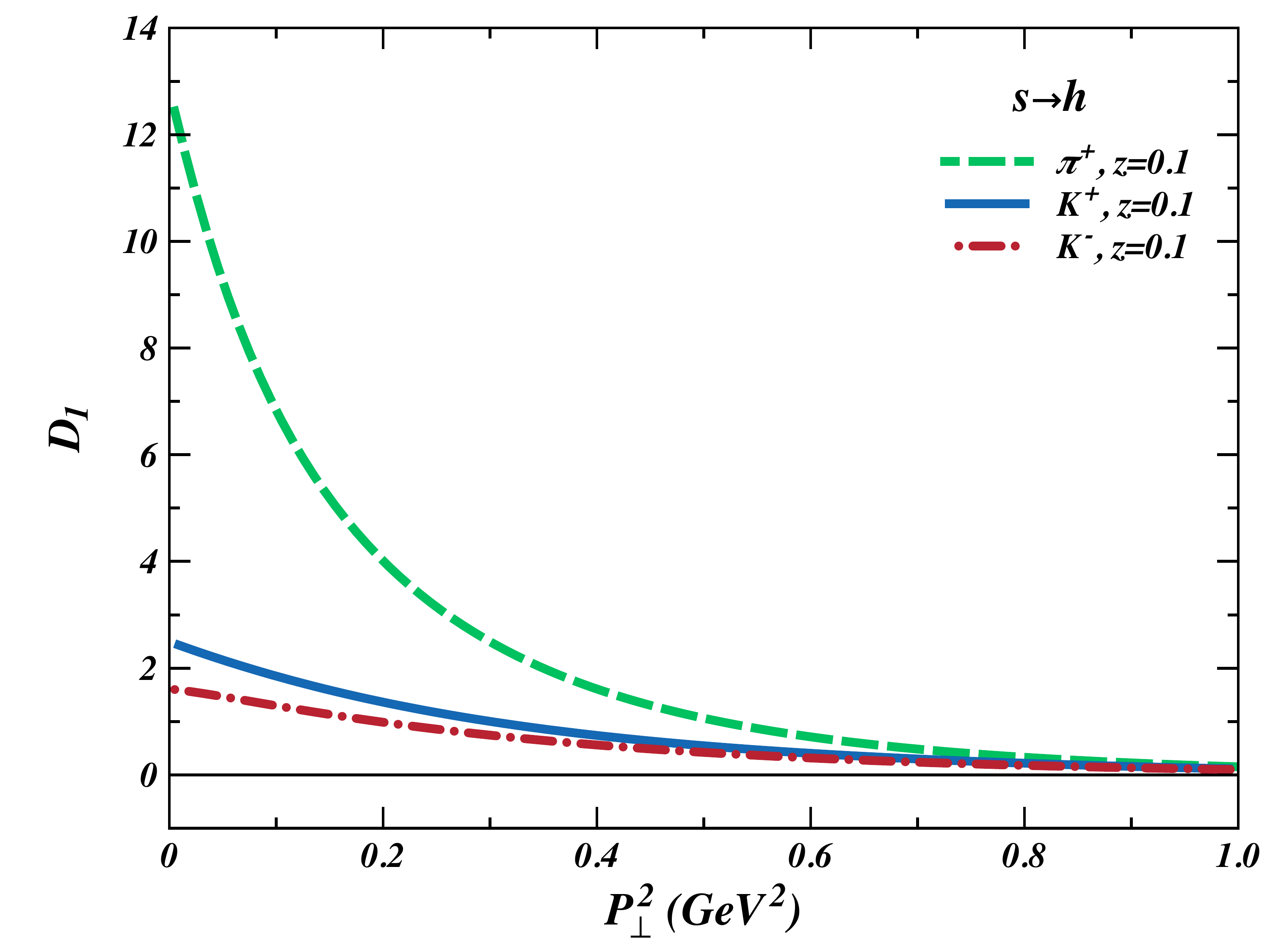}
}
\caption{The unpolarized fragmentation function vs $P_\perp^2$ for pions and kaons produced by a $s$ quark with $z=0.7$  (a)  and $z=0.1$ (b) for $N_{Links}=6$.}
\label{PLOT_TMD_D1_S}
\end{figure}

\begin{figure}[tbph]
\centering 
\subfigure[] {
\includegraphics[width=0.9\columnwidth]{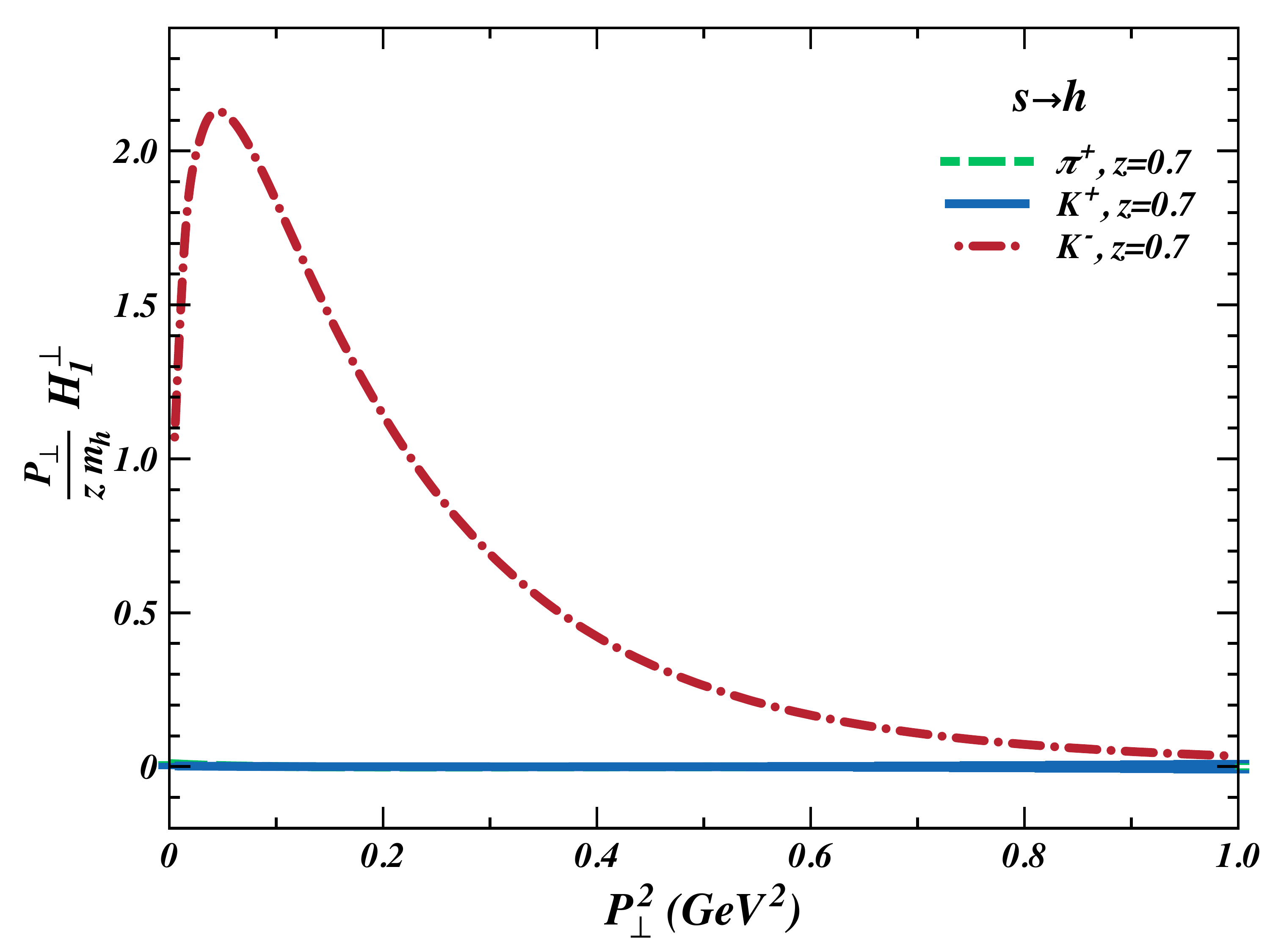}
}
\hspace{0cm} 
\subfigure[] {
\includegraphics[width=0.9\columnwidth]{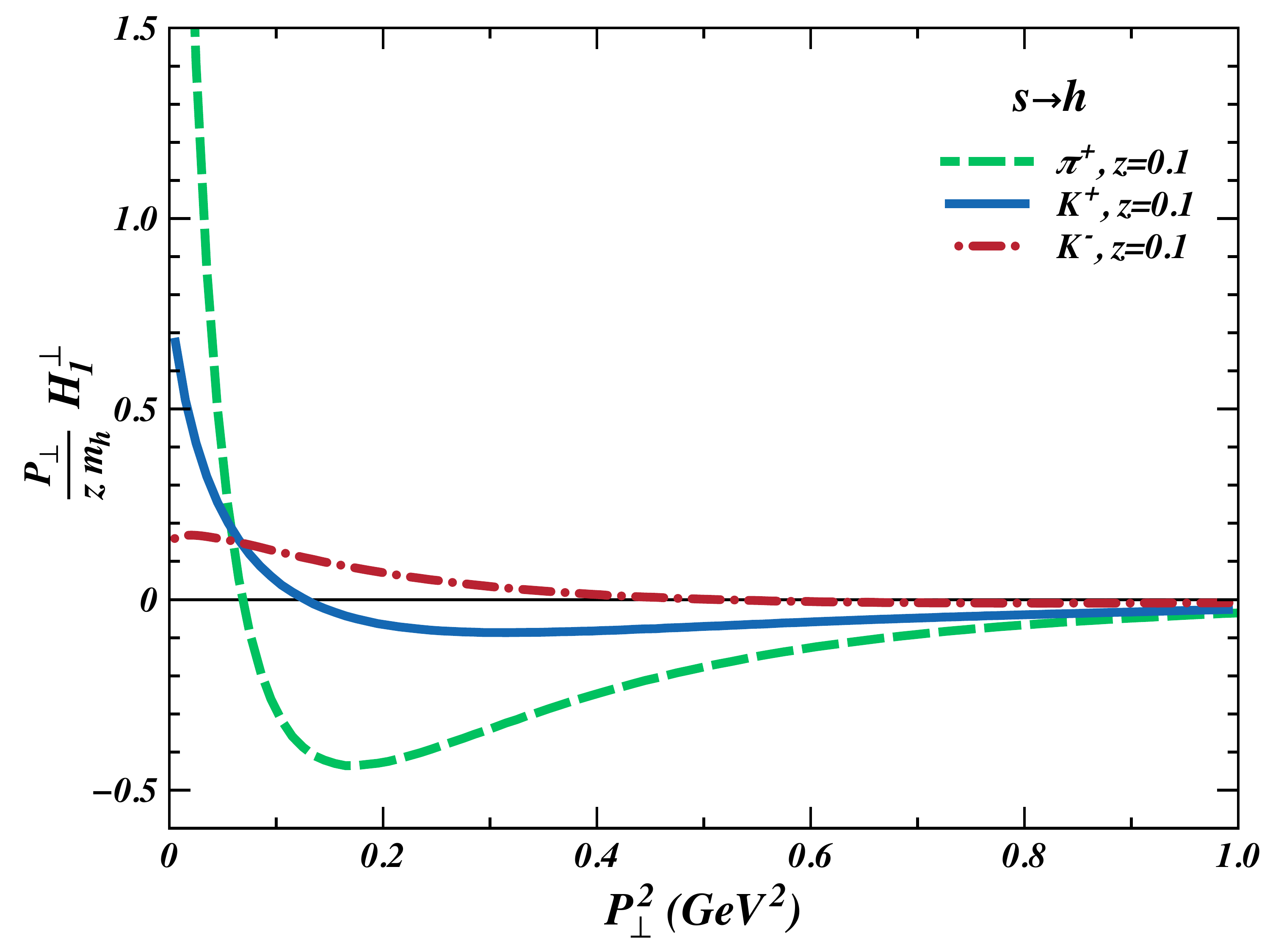}
}
\caption{The Collins fragmentation function vs $P_\perp^2$ for pions and kaons produced by a $s$ quark with $z=0.7$  (a)  and $z=0.1$ (b) for $N_{Links}=6$.}
\label{PLOT_TMD_COL_S}
\end{figure}

\begin{figure}[tbph]
\centering 
\subfigure[] {
\includegraphics[width=0.9\columnwidth]{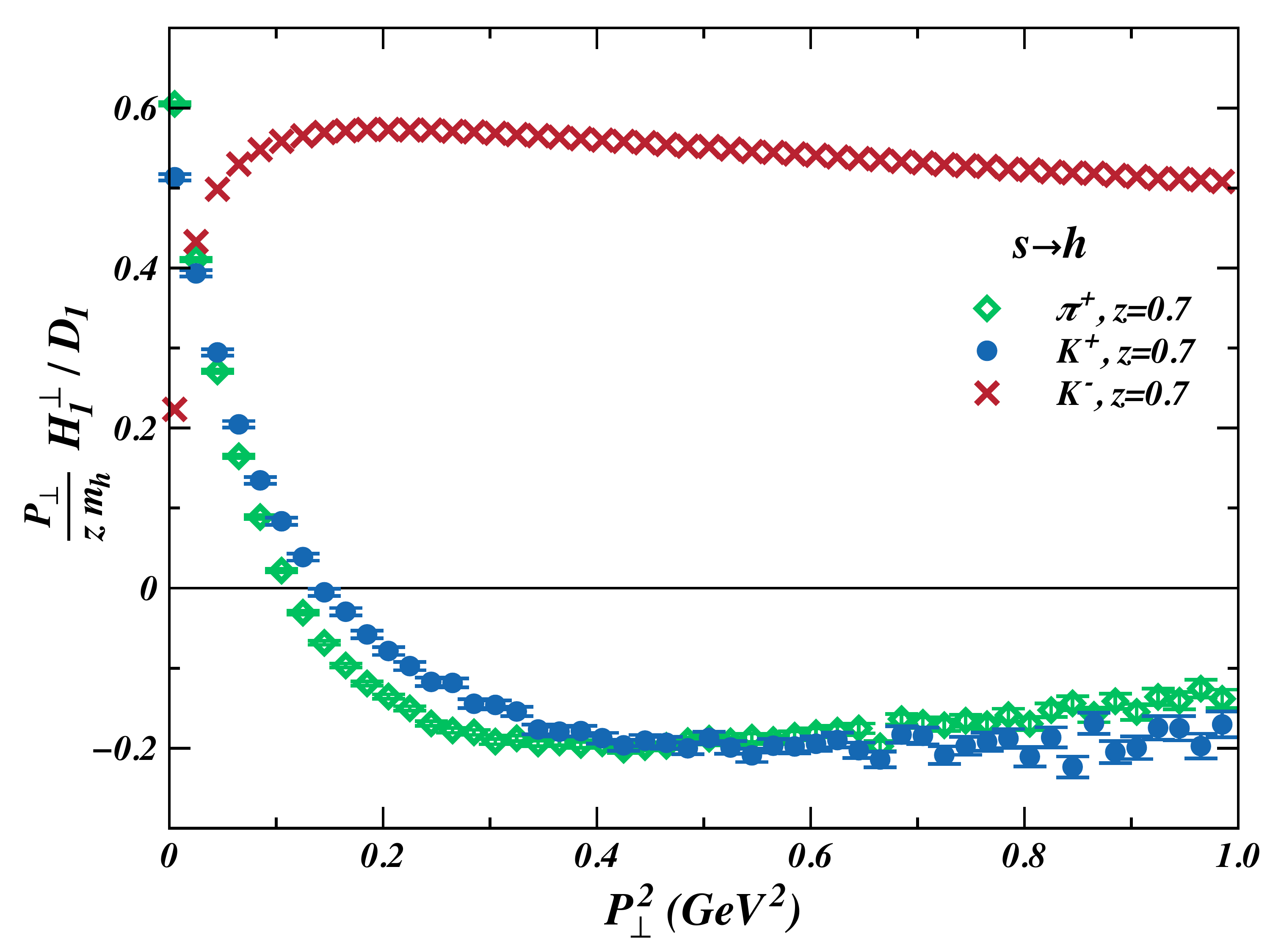}
}
\hspace{0cm} 
\subfigure[] {
\includegraphics[width=0.9\columnwidth]{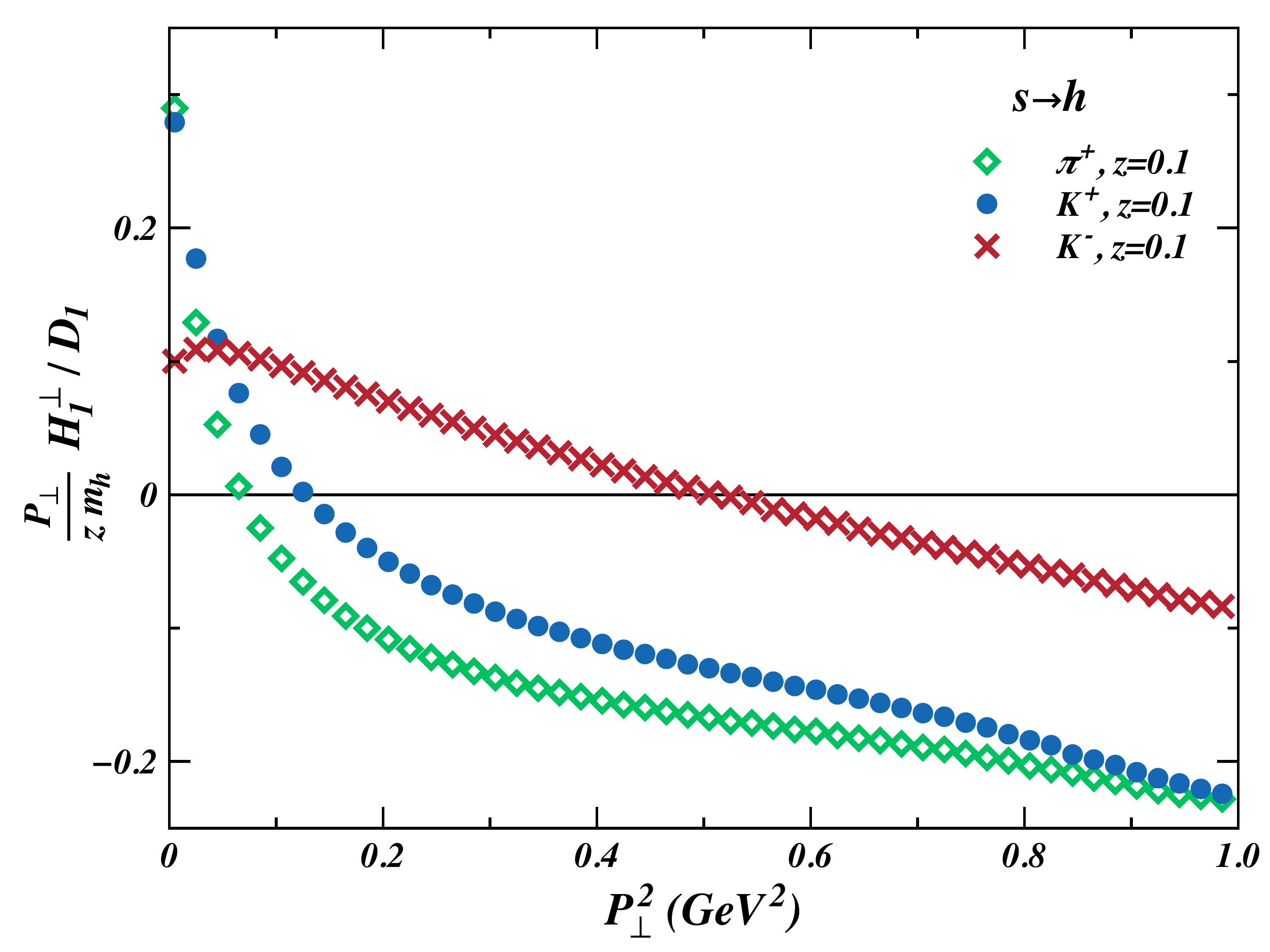}
}
\caption{The ratio of  Collins fragmentation function to unpolarized fragmentation vs $P_\perp^2$ for pions and kaons produced by a $s$ quark with $z=0.7$  (a)  and $z=0.1$ (b) for $N_{Links}=6$.}
\label{PLOT_TMD_RATCOL_S}
\end{figure}

The plots in Figs.~\ref{PLOT_TMD_D1_S},~\ref{PLOT_TMD_COL_S} and~\ref{PLOT_TMD_RATCOL_S} depict the results for unpolarized, Collins fragmentation functions and their ratios respectively for pions  and kaons produced by an initial $s$ quark with two fixed values of $z$ equal to $0.7$ and $0.1$. In our model the isospin symmetry is considered to be exact, thus the calculated values for all the pions are equal to each other, thus only $\pi^+$ is depicted. Similarly, the results for the kaons that can be transformed to each other by an isospin rotation coincide and are omitted as well.  The results are similar to those for the $u$ quark, except for the strong dominance of the favored fragmentation functions at $z=0.7$. At $z=0.1$, where multiple hadron emission contributions dominate, the unfavored fragmentation functions become larger than the favored ones, in particular for the production of pions.

\subsection{The Sch\"afer-Teryaev sum rule} 
\label{SEC_SUB_SUM_RULE}

  The Sch\"afer-Teryaev sum rule was originally proposed in Ref.~\cite{Schafer:1999kn}, where the authors used the simple arguments of the transverse momentum conservation in the hadronization process to derive
\begin{align}
\label{EQ_ST_SUM_RULE}
ST_q\equiv \sum_h \int_0^1 dz \ H_{1, (h/q)}^{\perp (1)}(z)=0,
\end{align}
where the first moment of the Collins function is defined as
\begin{align}
\label{EQ_COL_FIRST_MOM}
H_{1, (h/q)}^{\perp (1)}(z)\equiv \pi \int_0^\infty d P_\perp^2\  \frac{P_\perp^2}{2 z m_h} H_1^{\perp h/q}(z,P_\perp^2).
\end{align}

  Later, in Ref.~\cite{Meissner:2010cc}, it was proven explicitly using the quark correlator functions, along with the quark to hadron Collins function moments that the quark-to-quark Collins functions should also be included for the sum rule to be satisfied
\begin{align}
\label{EQ_ST_SUM_RULE_MEIS}
\sum_h \int dz \ H_{1, (h/q)}^{\perp (1)}(z)+\sum_Q \int dz \ 2 H_{1, (Q/q)}^{\perp (1)}(z)=0.
\end{align}

Our model results both for the toy model and the full calculation, show that the naive  Sch\"afer-Teryaev sum rule cannot be satisfied, even though the transverse momentum conservation is explicitly satisfied in our simulations. For example, the sum of the terms in Eq.~(\ref{EQ_ST_SUM_RULE}) for $u$ quark is $ST_u=0.07$, while for $s$ quark $ST_s= 0.21$.
 
 A simple explanation for this is that with multiple hadron emissions the quark quickly loses almost all of its initial light-cone momentum, but gains a nonzero average transverse momentum from the recoil of the emitted hadrons. This can be easily seen from the fact that the solutions of unpolarized integrated fragmentation functions, $D_1(z)$,  change only in the extremely small region of $z$, after only a few hadron emissions, as described in Ref.~\cite{Matevosyan:2011ey}. On the other hand, the average transverse momentum of the emitted hadrons tends to level off at a certain nonzero value as $z\to0$, as shown in Fig.~14 of \cite{Matevosyan:2011vj}, hinting at a similar value for the average transverse momentum of the remnant quark. In the NJL-jet model we assume that the remnant quark hadronizes with the slow, colored fragments of the initial struck hadron in SIDIS, or the remnant of the antiquark in $e^+e^-$ reactions, with the product unobserved because of the small values of remaining $z$ it carries. Thus, while the effects of the remnant quark after multiple hadron emissions can be neglected when considering the longitudinal momentum sum rules for the unpolarized fragmentations, they are essential for the transverse momentum sum rule for the Collins function, as shown in~\cite{Meissner:2010cc}.

\section{Conclusions}
\label{SEC_CONCLUSIONS}
\vspace{-0.2cm}

In this article we calculated the Collins fragmentation function in the NJL-jet model. This was accomplished by extending the model to include the transverse polarization of the fragmenting quarks and calculating the probability of the quark spin flip at each hadron emission. Then, we extended our Monte Carlo framework to accommodate for the spin of the quarks in the jet, and polar angle of the produced hadrons' transverse momenta with respect to the direction of the initial quark's spin. The polar-angle dependent elementary polarized fragmentations functions, that are a sum of elementary unpolarized and Collins functions, were used as inputs for the MC simulations. Here the unpolarized function was taken the one used earlier in NJL-jet model, while the Collins function was taken from calculations within the spectator model of~\cite{Bacchetta:2007wc}, with the parameters and regularization of transverse momentum integrals of the NJL-jet model. Using this input, we calculated the corresponding number densities for the produced hadrons in MC simulations, where we fixed the number of the produced hadrons in each quark-jet ($N_{Links}$). Then, we used the form of the polar angle dependence of the full number densities of Eq.~(\ref{EQ_Nqh}) to separate the unpolarized and the Collins functions.

 Results for two scenarios were presented. First, we showed the results for $P_\perp^2$ integrated case using a toy model, where we assumed the form of the elementary Collins term to be that of the unpolarized one times $0.1$. Also, the simulations were performed with only light quarks and pions for simplicity. The results for the solutions of $H_{1}^{\perp (1/2)}(z)$ showed that a nonzero unfavored Collins function can be generated in our model, with a sign opposite to that of the favored one. Also, both the unfavored and favored Collins functions oscillate for the lower values of $z$. Second, we presented the results with the full model calculations of both unpolarized and Collins functions with light and strange quarks, as well as pions and kaons. The resulting full solutions of the $1/2$ moment of the Collins functions exhibit features similar to the ones from the toy model. Interestingly, for the solutions for the kaons produced by a $u$ quark, the magnitudes of the unfavored solutions were only slightly smaller than the favored ones.
  
Finally, we studied the $P_\perp^2$ dependence of the Collins function for several fixed values of $z$.  Here we noticed that the multiple hadron emissions produce modulations of the polarized number densities, which contain several powers of $\sin(\varphi)$. These higher order modulations are a genuine feature of the quark-jet hadronization mechanism and neither depend on the particular forms for the elementary fragmentation functions nor are they artifacts of our Monte Carlo simulations. A detailed study, which will be presented in our forthcoming publications~\cite{Matevosyan:2012ms,Matevosyan:2012ed}, show that this effect can be completely described by the azimuthal modulation of the transverse momentum distribution of the remnant quark in each elementary fragmentation process and to the azimuthal modulations arising from the quark spin flip probability. The fitting functions for obtaining the unpolarized and Collins functions, described in Sec.~\ref{SEC_TOY_MODEL}, were modified using Eq.~(\ref{EQ_COL_GEN}) to account for this effect, which shows up only in the unintegrated TMD fragmentation functions. Our results show that the Gaussian function cannot be used to reliably model the Collins functions in any regions of $z$.

We also investigated the Sch\"afer-Teryaev sum rule for our results, and found that the na\"ive sum rule of Ref.~\cite{Schafer:1999kn} cannot be satisfied with our solutions, even though the transverse momentum conservation is explicitly enforced in our MC framework. The omission of the remnant quark transverse momentum is the key here, as it was proven explicitly from the definition of the Collins function in Ref.~\cite{Meissner:2010cc} that the sum rule holds if the quark-to-quark Collins functions are also accounted for. We argued that the remnant quark in the hadronization process after several hadron emissions, though carrying a minuscule fraction of the initial quark's light-cone momentum, acquires a significant transverse momentum on average, thus needs to be included in this sum rule that manifests the transverse momentum conservation in the hadronization process.

 Our results for Collins functions have some distinctive features that were approximately observed in the experiment: the similar size and the opposite sign for the $1/2$ moments of favored and unfavored ones. At this stage, however, a direct comparison with the experimental data is not possible because we did not perform the QCD evolution on our results, as these equations for the Collins function are not yet known. The recent work on the evolution of TMD distribution and fragmentation functions of Ref.~\cite{Aybat:2011zv,*Aybat:2011ge} pave the way for this, and we plan to implement the QCD evolution of Collins functions in our forthcoming work as the relevant formalism is developed.
  
\section*{Acknowledgements}

This work was supported by the Australian Research Council through Grants No. FL0992247 
(AWT), No. CE110001004 (CoEPP), and by the University of Adelaide. 

\bibliographystyle{apsrev}
\bibliography{}

\end{document}